\documentclass{ar-1col}

\usepackage{amsmath}
\usepackage{amssymb}
\usepackage{amsthm}
\usepackage{caption}
\usepackage{cleveref}
\usepackage{comment}
\usepackage{mwe}
\usepackage[comma]{natbib}
\usepackage{placeins}
\usepackage{subcaption}
\usepackage{url}
\usepackage{xcolor}

\Crefname{paragraph}{Section}{Sections}

\newtheorem*{theorem*}{Theorem}
\newtheorem*{example*}{Example}

\jname{Annual Review of Statistics and Its Application}
\jvol{9}
\jyear{2022}
\doi{10.1146/annurev-statistics-040220-091727}

\usepackage{graphicx}

\begin{document}

\markboth{South et al.}{Post-Processing of MCMC}

\title{Post-Processing of MCMC}

\author{Leah F. South,$^1$, Marina Riabiz,$^{2,3}$, Onur Teymur,$^{4,3}$ and Chris. J. Oates,$^{4,3}$
\affil{$^1$School of Mathematical Sciences, Queensland University of Technology, Brisbane, QLD 4000, Australia; email: 
l1.south@qut.edu.au}
\affil{$^2$Department of Biomedical Engineering, King's College London, SE1 7EH, UK}
\affil{$^3$Alan Turing Institute, London, NW1 2DB, UK}
\affil{$^4$School of Mathematics, Statistics \& Physics, Newcastle University, NE1 7RU, UK}}

\begin{abstract}
Markov chain Monte Carlo is the engine of modern Bayesian statistics, being used to approximate the posterior and derived quantities of interest.
Despite this, the issue of how the output from a Markov chain is post-processed and reported is often overlooked.
Convergence diagnostics can be used to control bias via burn-in removal, but these do not account for (common) situations where a limited computational budget engenders a bias-variance trade-off.
The aim of this article is to review state-of-the-art techniques for post-processing Markov chain output.
Our review covers methods based on discrepancy minimisation, which directly address the bias-variance trade-off, as well as general-purpose control variate methods for approximating expected quantities of interest.
\end{abstract}

\maketitle


\section{INTRODUCTION} \label{sec:introduction}

The Bayesian statistical framework is \textit{operational}, in the sense that a user first elicits their \textit{a priori} belief and then updates their belief in light of data, in a way that is (at least in principle) uniquely prescribed.
This updating is codified by \textit{Bayes' rule}, which expresses parameters \textit{posterior} probability density as being proportional to the product of \textit{a priori} probability density and the data likelihood.
Certain combinations of \textit{a priori} belief and likelihood are \textit{conjugate}, meaning that the posterior can be analytically computed.
Outside of the conjugate setting, computational methods are required.
The computational challenge, then, is to accurately approximate an \textit{intractable} probability distribution, meaning a distribution whose density function is available up to proportionality, where the normalisation constant is an intractable integral.

\begin{marginnote}[]
\entry{Intractable distribution}{A probability distribution whose density function is provided up to an unknown proportionality constant.}
\end{marginnote}

The majority of Bayesian analyses produce a posterior that is intractable, as indeed do other statistical frameworks \citep[such as generalised Bayesian inference;][]{bissiri2016general}.
There has, accordingly, been extensive research into computational methods for approximating intractable distributions.
The focus of this review is on \textit{Markov chain Monte Carlo} (MCMC) methods, a large class of computational methods which, for several decades now, have been considered among the state-of-the-art.
Given an intractable distribution, one can typically find several methods in the MCMC literature that can be applied.
However, the effectiveness of a particular method is not always easy to predict.
Furthermore, once an MCMC method has been applied, it is not always easy to determine the quality of the approximation produced.
Typical situations where these challenges occur include applications of Bayesian statistics in which the parameter space is high-dimensional, the likelihood has high information content, causing the posterior to present multiple modes or concentrate on manifolds, 
and settings where computational complexity limits the number of evaluations of the likelihood \citep{brooks2011handbook}.

Post-processing procedures aim to improve the quality of estimators that are based on MCMC output, either to approximate the probability distribution itself or a derived quantity of interest.
The main practical requirement of a post-processing procedure is that it should be agnostic to the details of the MCMC method.
The best known examples of post-processing procedures are \textit{burn-in removal} and \textit{thinning}.
In burn-in removal, one attempts to identify a number of iterations after which the Markov chain can (informally) be said to have \textit{converged} to the parameters posterior distribution, and then removes the initial part of the output where the chain had not converged.
This procedure can reduce bias in the MCMC output by reducing the dependence on how the MCMC was initialised, but it does not consider variance of the resulting estimators, which depends on the sample size and thus may be large if most of the chain is removed.
In thinning, every $k$th iteration is retained and the remainder discarded, in order to reduce the positive correlation between the remaining states, and therefore reduce the asymptotic variance of the estimators.
This can facilitate compression of MCMC output but does not always improve the approximation that is produced (notoriously, thinning does not lead to an efficiency gain if the samples are only used to estimate the posterior expectation of an inexpensive function).

It is thus notable that post-processing of MCMC engenders a bias-variance trade-off and yet standard post-processing procedures do not attempt to address this trade-off. 

This review focuses on modern post-processing techniques that can be applied to MCMC output.
Our discussion focuses on MCMC, for which consistency results have been established, but much of what we discuss is 
amenable to application in other computational methods that produce a collection of representative values as output, such as sequential Monte Carlo \citep{chopin2002sequential}.
To limit scope, our focus is principally on continuous-valued random variables, rather than discrete or categorical variables, but where possible we aim to keep discussion general.
The paper is structured as follows:
notation is established in \Cref{subsec: notation}, background on Markov chains is recalled in \Cref{subsec: markov chains}, and a formal problem statement is provided in \Cref{subsec: set up}.
\Cref{sec: approx posterior} focuses on the task of approximating the full probability distribution using MCMC output; we recall the standard approaches of burn-in removal and thinning, before describing modern and powerful techniques based on discrepancy minimisation in detail.
In \Cref{sec: approx expect} we focus on the task of approximating one or more scalar quantities of interest.
Control variate methods represent a powerful computational tool in this context, and we discuss the state-of-the-art in control variate methodology in detail.
A brief discussion concludes in \Cref{sec: discuss}.

\subsection{Notation}
\label{subsec: notation}

For this paper we use $(\Omega,\mathcal{F},\mathbb{P})$ to denote an underlying probability space on which all random variables are (often implicitly) defined and we let $\mathbb{E}[\; \cdot \;] = \int \cdot \; \mathrm{d}\mathbb{P}$.
Conditional probabilities are defined in the standard sense of \cite{kolmogorov1956foundations} and denoted $\mathbb{P}(F | G)$, $F,G \in \mathcal{F}$.
For this paper we introduce a measurable space $(\mathcal{X},\mathcal{B})$ and consider a random variable to be a measurable function $X : \Omega \rightarrow \mathcal{X}$, whose distribution $P$ is defined as $P(B) := \mathbb{P}(X \in B)$ for all $B \in \mathcal{B}$, where the conventional shorthand ``$X \in B$'' is used for the event $\{\omega \in \Omega : X(\omega) \in B\} \in \mathcal{F}$.
In the Bayesian context, $X$ represents the parameters of a statistical model and $P$ represents the posterior distribution after data have been assimilated.
Let $\mathcal{L}^2(P)$ be the vector space of random variables $f : \mathcal{X} \rightarrow \mathbb{R}$ with $\int f^2 \mathrm{d}P < \infty$. 
Let $\delta(x)$ denote the distribution of the random variable $f(\omega) = x$ for all $\omega \in \Omega$.
For a differentiable function $f : \mathbb{R}^d \rightarrow \mathbb{R}$, we denote the gradient of $f$ as $\nabla f$ where $(\nabla f)(x) := (\partial_{x_1}f(x), \dots , \partial_{x_d}f(x))^\top$.
For a differentiable function $f : \mathbb{R}^d \rightarrow \mathbb{R}^d$, we denote the divergence of $f$ as $\nabla \cdot f$ where $(\nabla \cdot f)(x) := \partial_{x_1}f_1(x) + \dots + \partial_{x_d}f_d(x)$, and if $f$ is twice-differentiable we denote the Laplacian of $f$ as $\Delta  f$ where $\Delta   f := \nabla \cdot (\nabla f)$. 
Natural numbers excluding zero are denoted $\mathbb{N}$ and including zero are denoted $\mathbb{N}_0$. 
The vector of ones is denoted $\mathbf{1}$, the unit vector $(1,0,\ldots,0)^{\top}$ is denoted $\mathbf{e}_1$ and $\|x\|$ denotes the Euclidean distance $\sqrt{x_1^2 + \dots + x_d^2}$.

\subsection{Markov Chains}
\label{subsec: markov chains}

A \textit{Markov chain} is a sequence $(X_n)_{n \in \mathbb{N}}$ of random variables $X_n : \Omega \rightarrow \mathcal{X}$ with the property that $X_{n+1} \perp\!\!\!\perp (X_m)_{m < n}\, |\, X_{n}$, where $X \perp\!\!\!\perp Y\, |\, Z$ indicates that the random variables $X$ and $Y$ are conditionally independent given the random variable $Z$.
\begin{marginnote}[]
\entry{Markov chain}{An ordered sequence of random variables $X_n$, such that $X_{n+1}$ is conditionally independent of $(X_m)_{m < n}$ given $X_n$.}
\end{marginnote}
In this paper we assume a non-random initial state $X_0 \in \mathcal{X}$.
To a Markov chain we can associate a sequence of \textit{transition kernels} $P_n(x,B) := \mathbb{P}(X_n \in B | X_{n-1} = x)$, $x \in \mathcal{X}$, $B \in \mathcal{B}$, so that $P_n(x,B)$ represents the probability that the state $X_n$ of the Markov chain takes a value in the set $B$, given that the previous state $X_{n-1}$ was equal to $x$.
The chain is said to be \textit{time-homogeneous} if $P_n$ does not depend on $n$.
Inductively define the \textit{$n$th step transition kernel} as $P^n(x,B) := \int P_n(y,B) P^{n-1}(x,\mathrm{d}y)$, $x \in \mathcal{X}$, $B \in \mathcal{B}$, with base case $P^0(x,B) = 1$ if $x \in B$, and $0$ if $x \notin B$.
That is, $P^n(x,B)$ represents the probability that the state $X_n$ of the Markov chain takes a value in the set $B$, given that the initial state $X_0$ was equal to $x$.
A Markov chain is said to be \textit{$P$-invariant} if $\int P^n(x,B) \mathrm{d}P(x) = P(B)$ for all $n$ and all $B \in \mathcal{B}$. 
Intuitively, if one was to randomise the initial state $X_0$ by sampling it from $P$, then the state $X_n$ will also have distribution $P$ if the Markov chain is $P$-invariant.
\begin{marginnote}
\entry{MCMC}{An MCMC method is an algorithm that, given a distribution $Q$, constructs a Markov chain that is $Q$-invariant.}
\end{marginnote}

Loosely speaking, a $P$-invariant Markov chain might be described as \textit{ergodic} if $P^n(x,B)$ approximates $P(B)$ in the $n \rightarrow \infty$ limit, for all $x \in \mathcal{X}$, $B \in \mathcal{B}$.
Several notions of ergodicity exist in the literature, but in this paper we focus on a specific notion called $V$-uniform ergodicity, which will now be defined.
For a function $V : \mathcal{X} \rightarrow [1,\infty)$, a function $f : \mathcal{X} \rightarrow \mathbb{R}$ and a measure $Q$ on $(\mathcal{X},\mathcal{B})$ we denote $\|f\|_V := \sup_{x \in \mathcal{X}} |f(x)| / V(x)$, $\|Q\|_V := \sup_{\|f\|_V \leq 1} |\int f \mathrm{d}Q|$.
A Markov chain is said to be \textit{$V$-uniformly ergodic} if there exist constants $R \in [0,\infty)$, $\rho \in [0,1)$, such that $\|P^n(x,\cdot) - P\|_V \leq R V(x) \rho^n$ for all $n \in \mathbb{N}$ and all $x \in \mathcal{X}$.
A comprehensive treatment of Markov chains can be found in the textbook of \cite{meyn2012markov}.

\subsection{Problem Statement}
\label{subsec: set up}

Consider an intractable probability distribution $P$.
Our aim is to compute an approximation, either to the distribution $P$ itself (\Cref{sec: approx posterior}) or to derived scalar quantities of interest (\Cref{sec: approx expect}).
Our starting point is one\footnote{In general applications it is common to exploit multi-core CPUs to simulate independent Markov chains in parallel. 
However, in the most challenging applications (where post-processing is most important), it is common to have access to only one MCMC output. 
This article considers post-processing of one MCMC output, but many of the methods we discuss can be trivially applied to aggregated MCMC output.} realisation (i.e. based on one random seed $\omega \sim \mathbb{P}$), of a finite portion $(X_n)_{n \leq N}$ of the Markov chain\footnote{To limit scope, so-called \emph{adaptive} MCMC, which aims to identify a suitable Markov transition kernel on-the-fly, will not be discussed. However, most of our presentation applies also to adaptive MCMC output.}, which we call the \textit{MCMC output}.  
It is \textit{not} assumed that the Markov chain is $P$-invariant unless stated, and later we discuss how output from a MCMC method that is $Q$-invariant may nevertheless enable $P$ to be consistently approximated if $Q$ is not too dissimilar to $P$.
All approximations are to be constructed by \textit{post-processing} the MCMC output.
In other words, we may only consider properties of $P$ defined locally at the states $X_n$ and no further exploration of $\mathcal{X}$ outside this finite set is permitted.
In particular, we exclude the trivial solutions of simply running further iterations of MCMC or adopting a different, possibly better MCMC method.
This set-up is realistic, reflecting the scenario that a practitioner has invested considerable resources into producing MCMC output and wishes to employ post-processing techniques to extract as much value as possible from their investment.

\begin{marginnote}[]
\entry{MCMC output}{A single realisation (or ``sample path'') of a Markov chain, of which the first $N$ states are provided.}
\end{marginnote}

An important preliminary comment is that the post-processing techniques described in this article (and indeed most MCMC methods) are \emph{not} parametrisation invariant.
This means that, if one were to apply an invertible transformation $Y_n = y(X_n)$, then post-processing of the MCMC output $(Y_n)_{n \leq N}$ can lead to different conclusions compared to if $(X_n)_{n \leq N}$ had been post-processed.
To limit scope we do not discuss parametrisation in this article.
Instead, following standard practice, we presuppose that one has employed transformation(s), such as centering and scaling \citep{yu2011center}, that (loosely speaking) promote simplicity, in order that $P$ can be more easily approximated.

\section{APPROXIMATION OF THE POSTERIOR DISTRIBUTION} \label{sec: approx posterior}

The outcome of an \textit{exploratory} Bayesian analysis is the posterior distribution itself, expressing \textit{a posteriori} belief about unknown parameters on the basis of \textit{a priori} belief and evidence provided by the dataset.
To facilitate exploratory Bayesian analysis outside the conjugate setting, it is therefore important that the entire posterior distribution can be accurately approximated. 
This section studies how MCMC output can be used to produce an approximation to a distribution $P$ of interest.
Throughout we consider approximations of the form 
\begin{equation}
\sum_{i=1}^M w_i \delta(X_{\pi(i)}), \label{eq: post processed}
\end{equation}
where $w_1,\dots,w_M \in \mathbb{R}$ are weights satisfying $\sum_{i=1}^M{w_i} = 1$ and $\pi:\{1,\dots,M\} \rightarrow \{1,\dots,N\}$ is a function that indicates which states from the MCMC output are included in \Cref{eq: post processed}.
In simple terms, this approximation extracts and re-weights a subsequence of length $M$ from the given MCMC output of length $N$. 

\begin{marginnote}[]
\entry{Post-processing MCMC output}{Selecting a weighted combination of states from the MCMC output to better represent the posterior distribution $P$.}
\end{marginnote}

Recall the two categories of post-processing discussed in Section \ref{sec:introduction}. 
First, if the chain is constructed so that its asymptotic law converges to $P$, then excluding the first $b$ points (the \emph{burn-in} period) from \Cref{eq: post processed} may help to reduce bias due to the choice of the initial state $X_0$ of the Markov chain. 
This corresponds to excluding $\{1,\dots,b\}$ from the image of $\pi$, and we discuss standard approaches to this problem in Section \ref{ssec:burnin}. 
Second, thinning of MCMC output can be useful when samples are to be used for further computation, especially when the subsequent computation has a high 
cost. 
This corresponds to excluding $i$ from the image of $\pi$ whenever $i \neq 1$ modulo $k$, and we briefly discuss approaches to thinning in Section \ref{ssec:mindiscrepancy}. 
In both cases, uniform weights $w_i = \frac{1}{M}$ are assumed in \Cref{eq: post processed}.

\subsection{Burn-in Removal} \label{ssec:burnin}

\begin{marginnote}[]
\entry{Burn-in}{The first $b$ states of a $P$-invariant Markov chain, for which the distribution of $X_n$, $n \leq b$, is deemed to substantially differ to $P$.}
\end{marginnote}
	
    In this section we discuss standard approaches to identification of a burn-in period from given MCMC output, in order to control the bias resulting from an arbitrary choice of initial state $X_0$ for the Markov chain.
    Our focus is limited to continuous domains $\mathcal{X} \subseteq \mathbb{R}^d$. 
	Rigorous approaches for selecting a burn-in period $b$ have been proposed by authors including \cite{meyn1994computable, rosenthal1995minorization, roberts1999bounds}; see also \cite{jones2001honest}.
	Unfortunately, these often involve conditions that are difficult to establish or, when they hold, they may provide loose bounds on the total variation distance between the law of the Markov chain and the invariant distribution, implying an unreasonably long burn-in period. 
	More recently 
	\cite{biswas2019estimating}  discuss how to estimate such bounds through coupling and multiple MCMC runs, but this is out of the scope considered here, where  a single MCMC run has been obtained at  moderate to high computing cost.  
	Convergence diagnostics have 
	emerged as a practical solution to the need to test for non-convergence of MCMC. 
	Their use is limited to reducing bias in MCMC output; they are not designed for the setting that we consider, where the length $N$ of the MCMC output is fixed, and which requires a bias-variance trade-off.
	Nevertheless, convergence diagnostics constitute the most common means by which MCMC output is post-processed in modern software packages for MCMC, including \verb+WinBUGS+ \citep{lunn2000winbugs}, \verb+JAGS+ \citep{plummer2003jags}, \verb+R+ \citep{Rproject},  \verb+Stan+ \citep{carpenter2017stan}, and \verb+PyMC3+ \citep{salvatier2016probabilistic}.

	In this section we recall standard practice for selection of a burn-in period $b$, and thus (implicitly) in constructing an estimator of the form \Cref{eq: post processed}, focussing on the traditional $\widehat{R}$ statistic of 
	\cite{gelman1992,brooks1998general}.
	The aim of this Section is to describe the general idea and fundamental limitations of convergence diagnostics in the fixed $N$ scenario, rather than presenting the state-of-the-art or providing a comprehensive survey of convergence diagnostics for burn-in removal. 
    We simply recall that the $\widehat{R}$ convergence diagnostic was first introduced in \cite{gelman1992} and subsequently corrected in \cite{brooks1998general}, and this was then simplified in \cite{gelman2003}. We use the implementation of the \cite{brooks1998general} version from the \texttt{R} package \texttt{coda} in our experiments and we focus on the simple expression of \cite{gelman2003} in the text. Further developments of the $\widehat{R}$ convergence diagnostic include 
    \cite{gelman2014}, where the diagnostic test is performed separately on each half of the MCMC output;  \cite{vats2018revisiting}, that revisits a connection between $\widehat{R}$ and effective sample size (ESS) of quantities of interest estimated from the MCMC output;\footnote{The ESS indicates how many independent samples are needed to provide the same amount of information, about that quantity of interest, as the correlated MCMC output: the higher is this value, the lower is the loss of information due to correlation in the MCMC output.}  \cite{vehtari2021rank}, that provides more details on such connections, addressing also target distributions with infinite variance, and the case in which the Markov chain is exploring the bulk of the target distribution, but not its tails. 
    A comprehensive survey of convergence diagnostics for MCMC can be found in \citet{Roy2020}.

	The traditional $\widehat{R}$ statistic of \cite{gelman2003} is not a post-processing method in the strict sense set out in \Cref{subsec: set up}, because it is based on $l = 1,\dots,L$ independent realisations of MCMC output, $(X_n^l)_{n \leq N}$; i.e. $X_n^l$ denotes the random variable $X_n$ evaluated at $\omega_l$, where $\omega_1,\dots,\omega_L \sim \mathbb{P}$ are independent.
	For a uni-dimensional target distribution, 
	the traditional $\widehat{R}$ statistic is defined as the  square root of the ratio of two estimators of the variance $\sigma^2$ of the target:
	\begin{align*}
	\widehat{R}
	:= \sqrt{\frac{\hat{\sigma}^2}{s^2}},
	\label{eq:GR_diagnostic}
	\end{align*}
	where $s^2$ is the (arithmetic) mean of the sample variances $s^2_l$ along the $L$ sample paths $(X_n^l)_{n\leq N}$, and it  typically provides an underestimate of $\sigma^2$; 
	while 
	$\hat{\sigma}^2$ is constructed as an unbiased overestimate of the target variance
	\begin{align*}
	\hat{\sigma}^2 := \frac{N-1}{N} s^2 + \frac{1}{L-1} \sum_{l=1}^L \left( m_{l} - \frac{1}{L} \sum_{l'=1}^L m_l \right)^2 ,    
	\end{align*}
    where $m_l$ is the sample mean from the $l$th sample path, and where the second term is the sample variance of the sample means from $L$ chains. 
	\begin{marginnote}[]
    \entry{Sample mean and variance}{The sample mean $m_l$ and variance $s_l^2$ of a MCMC output $(X_n^l)_{n \leq N}$ are defined, respectively, as $\frac{1}{N} \sum X_n^l$ and  $\frac{1}{(N-1)} \sum (X_n^l -m_l)^2 $, where the sums run over $n = 1,\dots,N$.}
    \end{marginnote}
	For an ergodic Markov chain, 
	$\widehat{R}$ 	
	converges to 1 as $N \rightarrow\infty$.
	In practice, it is common to discard a burn-in period of length $b = N$, where $N$ is the smallest integer for which $\widehat{R} < 1 + \delta$, and $\delta$ is a suitable threshold\footnote{Although seeking the \emph{smallest} $N$ is not an explicit recommendation in the literature cited, it is clear that one would not want to simulate a Markov chain for longer than required. Thus, in effect, it is standard practice to work with $N$ as small as possible, subject to the diagnostic test being passed.}. 
	The somewhat arbitrary choice of $\delta = 0.1$ has historically been used \citep{gelman2014}, and current best practice for traditional $\widehat{R}$ and its extensions advocates $\delta = 0.01$ \citep{vehtari2021rank}.

\begin{figure}[t!]
    \centering
    \includegraphics[width=0.7\textwidth]{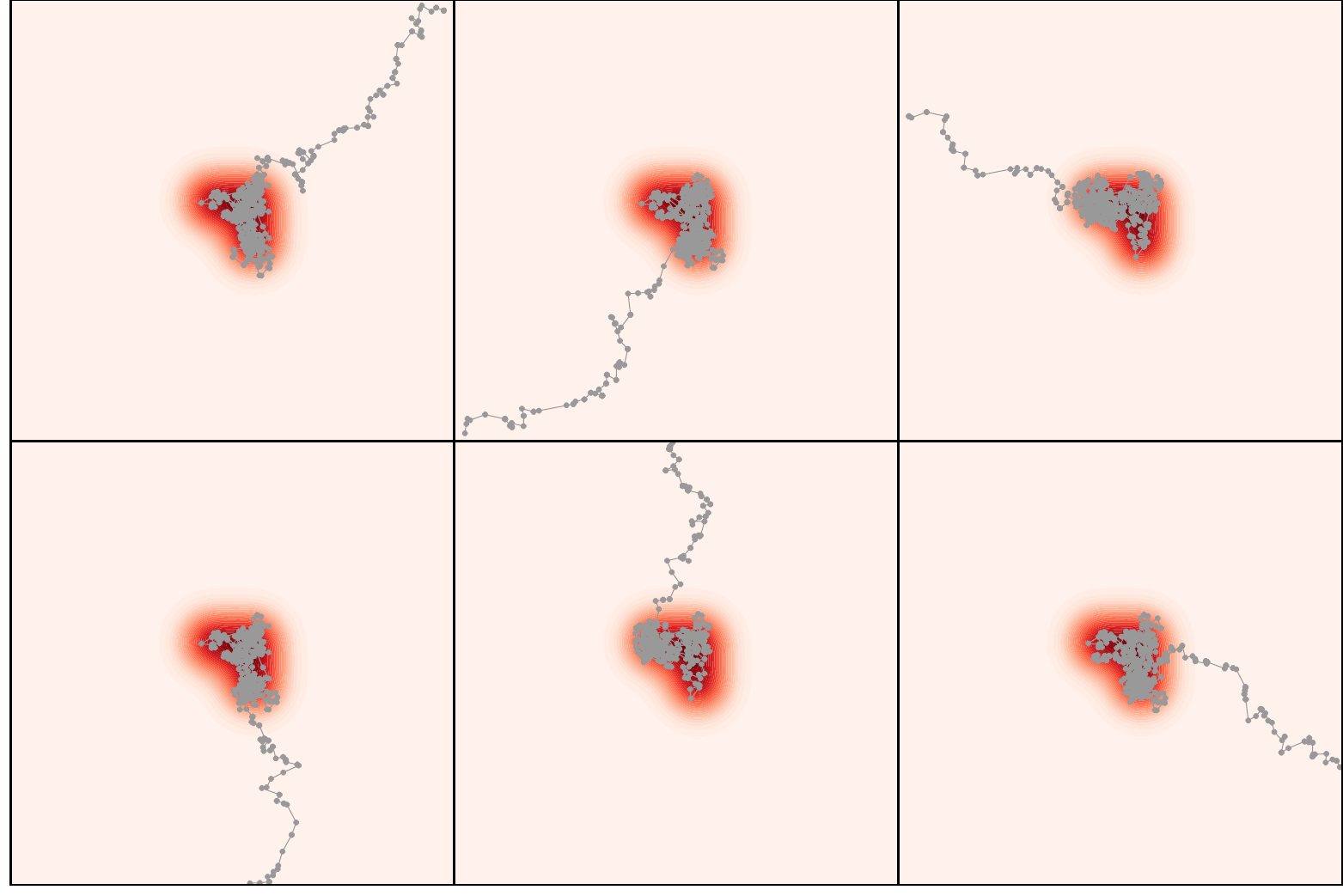}
   \captionof{figure}{MCMC output. Here we show $L=6$ independent realisations of MCMC output (gray lines), for a particular bivariate distributional target $P$ indicated by the shaded contour plot in the background.
   In each case a total of $N = 10^3$ iterations of the Markov chain were performed, with the first $500$ iterations plotted.
   }
   \label{fig:MCMC_outputs}
\end{figure}

\begin{figure}[t!]
    \centering
    \includegraphics[width=1\textwidth]{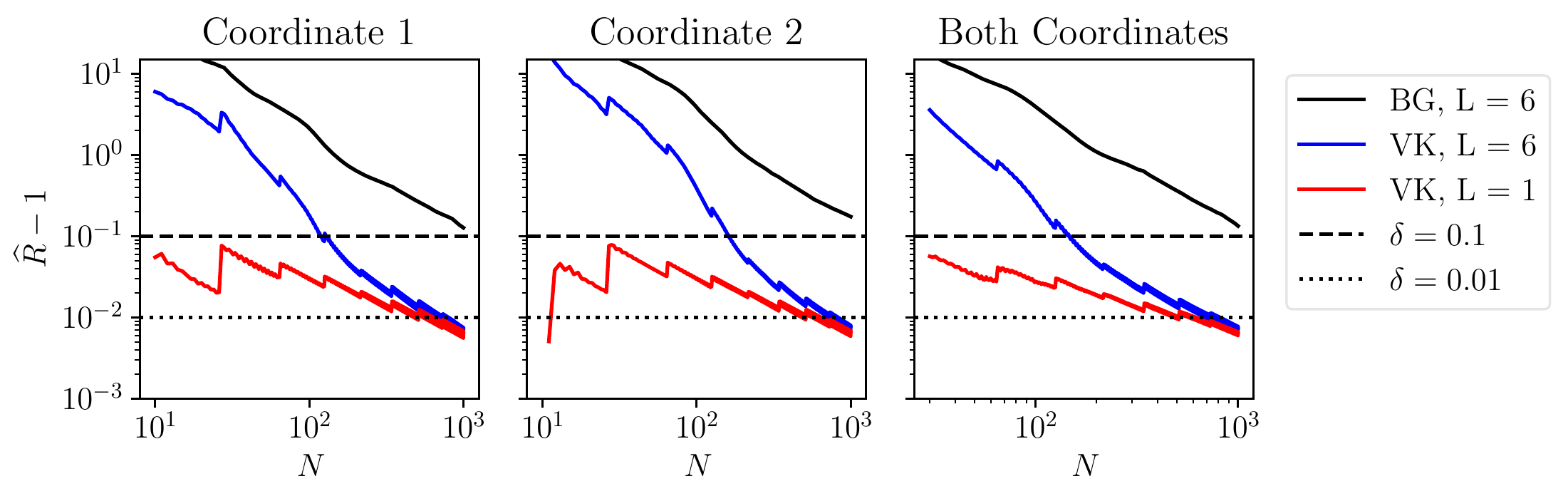}
   \captionof{figure}{Convergence diagnostics for the MCMC output shown in \Cref{fig:MCMC_outputs}. 
   Here we show the traditional $\widehat{R}$ statistic of \cite{brooks1998general} (BG; 
   black solid lines) 
   and also an autocorrelation based diagnostic used by \cite{vats2018revisiting} (VR; blue and red solid lines), each as a function of the number of iterations $N$ of the Markov chain that are considered. 
   These diagnostics were applied separately to the first and second coordinates of the bivariate state variable (left and central panels) and jointly to both coordinates (right panel).
   Dashed lines indicate thresholds at which convergence is deemed to have occurred. In all cases, 
   the traditional $\widehat{R}$ statistic 
    does not 
   fall below the  thresholds, indicating that convergence has \emph{not} occurred.
   }
   \label{fig:R_hat}
\end{figure}

Convergence diagnostics can help to detect situations in which a Markov chain has not converged, and for this purpose they are widely used.
Their main drawbacks are (a) such diagnostics do not provide guarantees that the Markov chain has actually converged (existing convergence diagnostics can assess only necessary but not sufficient conditions for convergence); (b) burn-in removal may not be useful in practical settings where the MCMC output has already been obtained and post-processing is required, as described in Section \ref{subsec: set up}. 
In order to mitigate the first point,  \cite{vats2018revisiting, vehtari2021rank} recommend to look at the effective sample size of quantities of interest, if possible combining autocorrelation information from multiple chains, which helps detecting poor convergence in cases of multimodal target distributions. 
However, this still remains only a necessary, not sufficient condition for convergence, and it does not help tackling the second point. 
This section ends with an example to highlight this important second drawback of convergence diagnostics:

\begin{example*}[Burn-in removal lacks a bias-variance trade-off]
The purpose of convergence diagnostics is to detect and avoid bias due to dependence on the arbitrary choice of initial state $X_0$.
However, burn-in removal does not address the bias-variance trade-off that occurs when the MCMC output is fixed.
As an extreme illustration of this, consider the MCMC output shown in \Cref{fig:MCMC_outputs}. 
Here $L=6$ independent sample paths of total length $N = 10^3$ were produced 
using random walk Metropolis--Hastings \citep{Metropolis1953}. 
A simple bivariate target $P$, whose contour lines are plotted in red, was used, but the Markov chain was not optimised, to simulate a challenging sampling context. 
The initial states $X_0^l$ were over-dispersed relative to the target $P$, requiring the Markov chain to take several steps before the high probability region is reached. 
\Cref{fig:R_hat} applies convergence diagnostics to establish whether or not the Markov chains can be said to have converged.

The traditional $\widehat{R}$ statistic 
of \cite{brooks1998general} 
(black solid line) detects non-convergence even after all $N = 10^3$ iterations of the MCMC output have been considered, irrespective of whether the diagnostics are applied to each coordinate of the state vector or jointly to both coordinates.
This is undesirable from our perspective of post-processing MCMC output, since it is clear from \Cref{fig:MCMC_outputs} that there is useful information in the MCMC output, even if 
some 
dependence on $X_0$ can 
be detected.
In addition, we present in \Cref{fig:R_hat}  two of many proposed improvements over \citet{brooks1998general}: 
the recent diagnostic due to \cite{vats2018revisiting} (blue solid line), and also 
a version of such convergence diagnostic (red solid lines), presented in the same work, and that can be computed using a single MCMC output. 
They indicate that the burn-in period has finished, but they 
leave only a small portion of the chain after the burn-in, when considering the threshold $\delta = 0.01$. 
All convergence diagnostics were computed using the {\normalfont \texttt{R}} packages {\normalfont \texttt{coda}} \citep{Rcoda} and {\normalfont \texttt{stableGR}} \citep{RstableGR}.

The modern MCMC post-processing techniques presented in Section \ref{ssec:mindiscrepancy}  address this bias-variance trade-off, and their use is encouraged in  problems where obtaining further MCMC iterations is not practical.
\end{example*}

\subsection{Fixed Frequency Thinning} \label{ssec:thinning}

As with the classical approaches to burn-in removal discussed in Section \ref{ssec:burnin}, thinning is often performed on a heuristic basis as the simplest way to achieve compression of MCMC output. 
In exploratory Bayesian analysis,  this is often motivated by the need to reduce storage cost or to make subsequent computation faster (the reader can, for example, think about the case in which the samples obtained from the posterior are used for forward uncertainty propagation in complex multi-scale models). However, thinning is traditionally performed with no specific aim to improve the accuracy of the MCMC output. 

The most common approach to thinning is to sub-sample with a fixed frequency from the chain (`retain every $k$th sample and discard the rest'), which can be an effective strategy to reduce auto-correlation in the MCMC output.
Systematic approaches for determining $k$ do exist, and 
the most well-known is based on the auto-correlation estimator of \cite{geyer92}, that can be computed using the \texttt{R} package \texttt{LaplacesDemon} \citep{LaplDemon21}. This method estimates a sequence of fixed-lag auto-correlations in the Markov chain, and then thresholds this sequence to give a $k$ that results in a subsample that is close to uncorrelated. This procedure is most useful in exploratory Bayesian analysis, where a set of such samples are themselves required, rather than as an attempt to improve an estimator. See \cite{owen2017} for a discussion on the statistical efficiency of this approach.

More sophisticated approaches to compress MCMC have been explored by authors including \cite{paige2016super} and \cite{mak2018support}.
In both cases, the authors aimed to construct an approximation of the form in \Cref{eq: post processed} with $M \ll N$, such that \Cref{eq: post processed} provides an accurate approximation to the discrete distribution $\frac{1}{N} \sum_{n=1}^N \delta(X_n)$ supported on the original MCMC output.
Although these can provide effective compression of MCMC output, if the Markov chain has not converged then the compressed output will be biased. 
In the next section, we discuss an approach that is simultaneously capable of thinning and de-biasing MCMC output, and is applicable even in cases where the Markov chain is not $P$-invariant.

\subsection{Discrepancy Minimisation}\label{ssec:mindiscrepancy}

Here we discuss modern and powerful approaches to post-processing of MCMC that aim to directly address the bias-variance trade-off just described.
The approach we will explore casts the choice of $\pi$ in \Cref{eq: post processed} as an optimisation problem. 
The key idea is to identify an appropriate quantification of the discrepancy between the discrete distribution $Q_M$ in \Cref{eq: post processed} and the distributional target $P$, and then to select both the weights $w_i$ and the index sequence $\pi$ so that this discrepancy is minimised. 
A \emph{discrepancy} is defined as a bivariate function $D$ such that $D(P,P) = 0$ for all distributions $P$, and $D(P,Q) > 0$ for all $P \neq Q$. 
There are an infinitude of functions $D$ that satisfy these relations, so for a discrepancy to be useful we typically require several other properties.
An important property, which we will not discuss further in this paper since it is often satisfied, is that $D(P,Q_M) \rightarrow 0$ whenever $Q_M$ converges to $P$ in an appropriate sense.
Another important property, which is the converse of the property just described and which we will discuss, is called \textit{convergence control}, where $D(P,Q_M) \rightarrow 0$ implies that $Q_M$ converges to $P$ in a sense that must be specified. 
\begin{marginnote}[]
\entry{Discrepancy}{A discrepancy $D$ is a non-negative function where $D(P,Q)$ is interpreted as the dissimilarity between measures $P$ and $Q$.}
\end{marginnote}

\begin{marginnote}[]
\entry{Convergence control}{A discrepancy $D$ is said to have convergence control if $D(P,Q_M) \rightarrow 0$ implies $Q_M$ converges to $P$ in a sense that must be specified.}
\end{marginnote}

As an aside, we note that many related strands of work exploit discrepancy to approximate a distributional target $P$ by a discrete distribution $Q^N$. 
For example, in quasi Monte Carlo an \textit{explicit} construction $Q^N = \frac{1}{N} \sum_{n=1}^N \delta(x_n)$ is sought to approximate $P$ in such a way that a discrepancy $D(P,Q^N)$ is provably asymptotically minimised \citep{hickernell1998generalized,dick2010digital}.
In a different direction, other researchers have \textit{implicitly} constructed point sets by performing direct minimisation of $D(P,Q^N)$ over the high-dimensional joint space $(x_1,\dots,x_N) \in \mathcal{X}^N$; see the surveys in \cite{Briol2017,oettershagen2017construction,pronzato2018bayesian,briol2019probabilistic}. 
In both of these methods the goal is to find a compressed representation of $P$, and as a starting point it is assumed that $P$ is known in full. This is not the case when one has to post-process MCMC output. An approach considered by \citet{Kyriazopoulou2008,Kontoyiannis2008} is to adjust estimates based on the discrepancy between the expectation under $P$ and the expectation under $Q^N$ of a reference function, although this discrepancy is not convergence-determining for a finite number of reference functions. To overcome these problems a specialised family of discrepancies will be required.
These are introduced next.

\begin{figure}[t!]
    \centering
   \includegraphics[width=0.4\textwidth,clip,trim = 0.5cm 0.6cm 0cm 0cm]{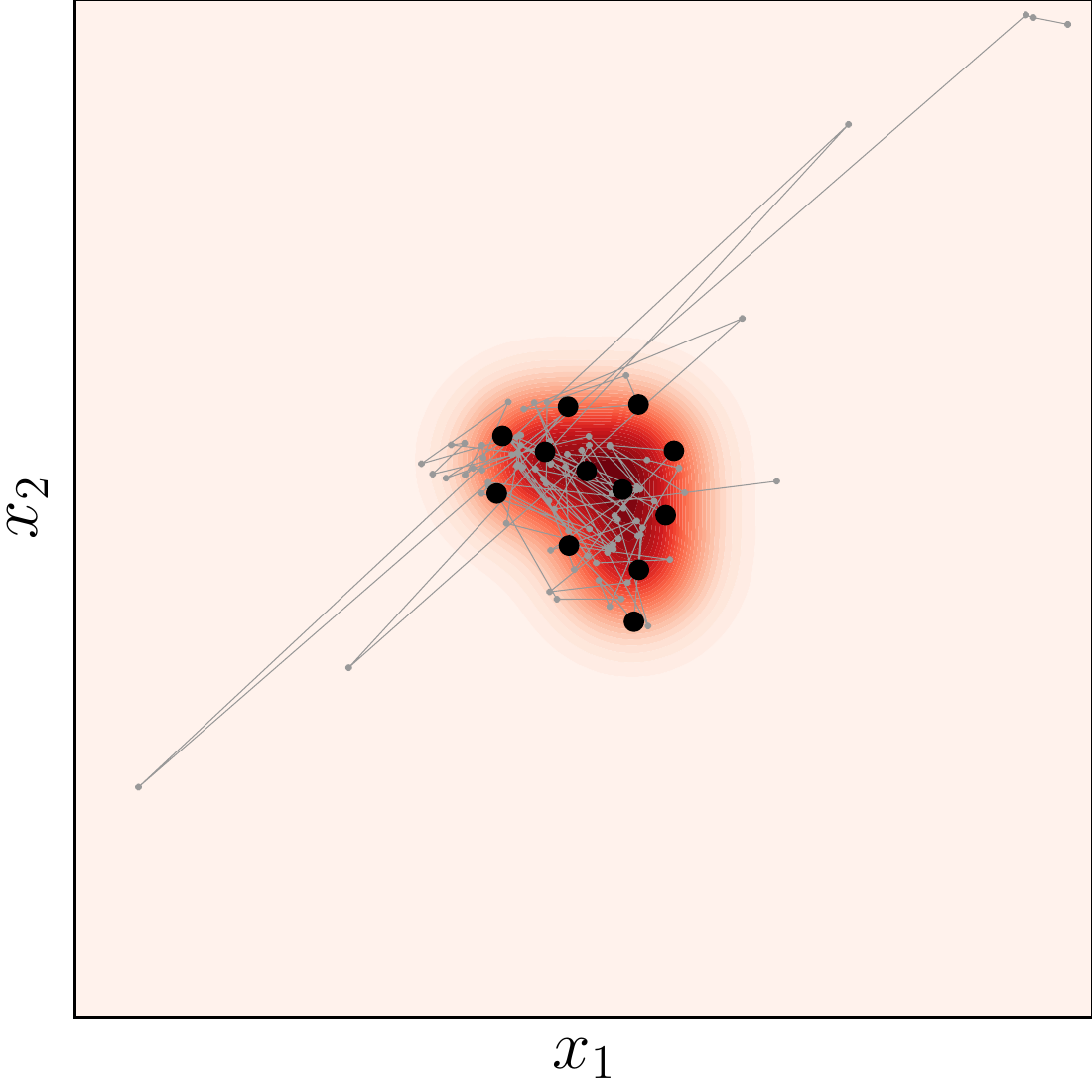}
   \captionof{figure}{Post-processing of MCMC output via Stein thinning. Here a Markov chain sample path (gray line) is post-processed to select $M=12$ representative states (black circles), such that the discrete measure supported on these representative states provides an accurate approximation to the same distributional target $P$ as in Figure \ref{fig:MCMC_outputs}, indicated by the shaded contour plot in the background. }
   \label{fig:MCMC_output_processed}
\end{figure}

\subsubsection{Stein Discrepancy Minimisation}
\label{ssec: stein disc min}

Our aim is to select an appropriate discrepancy $D$ for post-processing of MCMC. 
To this end, we focus on \emph{Stein discrepancy}, and in particular a \emph{kernel Stein discrepancy} constructed for the case where the domain $\mathcal{X}$ is $\mathbb{R}^d$; see the three inset boxes for definitions and detail. 
The main computational requirement when using Stein discrepancy is the evaluation of the gradient $\nabla \log p$ along the MCMC sample path, where $p$ is a density function for $P$.
Note that gradient-based samplers, such as the Metropolis-adjusted Langevin algorithm \citep{Roberts2002} or Hamiltonian Monte Carlo \citep{Duane1987}, produce the required evaluations as a by-product.
Stein discrepancy is particularly well-suited to post-processing of such MCMC output since, under appropriate technical assumptions, it (a) allows explicit computation of $D(P,Q_M)$, and (b) provides convergence control, meaning in this context that $D(P,Q_M)\rightarrow 0$ implies $Q_M$ converges weakly to $P$.

\begin{marginnote}[]
\entry{Stein Discrepancy}{A discrepancy $D$ such that $D(P,Q)$ can be computed when $P$ is an intractable distribution and $Q$ has a finite support.}
\end{marginnote}

\begin{textbox}[t!]
\section{STEIN DISCREPANCY}
A \textit{Stein discrepancy} is a discrepancy of the form
\begin{equation}
    D(P,Q) = \sup_{f \in \mathcal{F}_P} \left| \int f \mathrm{d}Q \right|, \label{eq: stein disc}
\end{equation}
where $\mathcal{F}_P$ is a set of functions chosen to satisfy $\int f \mathrm{d}P = 0$. 
For a sufficiently large set $\mathcal{F}_P$ it is possible to have $D(P,Q) = 0$ imply $P = Q$.
One way of achieving this is by taking $\mathcal{F}_P$ to be the set of functions of the form $f(x) = h(x) - \int h \mathrm{d}P$, with $h$ ranging over a measure-determining set $\mathcal{H}$ \citep[such a $D$ is recognised as an \textit{integral probability metric}; see][]{muller1997integral}.
However, for intractable $P$ the presence of the integral $\int h \mathrm{d}P$ renders this choice impractical.

Building on \cite{Stein1972}, the recent work of \cite{gorham2015measuring} proposed an alternative approach that can be applied provided that $P$ admits a positive and differentiable density on $\mathcal{X} = \mathbb{R}^d$.
Let the set $\mathcal{F}_P$ be composed of functions of the form 
\begin{align}
f(x) = (\mathcal{A}_P h)(x) := (\nabla\cdot h)(x) +  (\nabla \log p)(x) \cdot h(x) , \label{eq: Ap}    
\end{align}
where $h$ ranges over a sufficiently large set $\mathcal{H}$ of differentiable functions $h : \mathbb{R}^d \rightarrow \mathbb{R}^d$.
The differential operator $\mathcal{A}_P$ is called a \textit{Stein operator} and the set $\mathcal{F}_P$ is called a \textit{Stein set}.
Under a particular \emph{tail condition} on $h$ (see inset box) it can be shown that $\int f \mathrm{d}P = 0$.
With further regularity assumptions, it can be shown that such a Stein discrepancy can enjoy either Wasserstein convergence control \citep[see][Theorem 2]{gorham2015measuring} or weak convergence control \citep[see][Theorem 8]{gorham2017measuring}, depending on how the set $\mathcal{H}$ is selected.
Alternative Stein operators are also possible; see \cite{gorham2019}.
\end{textbox}

\begin{textbox}[t!]
\section{TAIL CONDITION FOR STEIN DISCREPANCY}
To construct a Stein discrepancy we require a set of functions $h : \mathbb{R}^d \rightarrow \mathbb{R}^d$ for which the \textit{Stein identity} $\int (\mathcal{A}_P h) \mathrm{d}P = 0$ holds, with $\mathcal{A}_P$ defined in \Cref{eq: Ap}.
This can be formulated as a \emph{tail condition} on $h$.
The main idea is to recognise $\mathcal{A}_P h$ as a divergence operator and exploit the \textit{divergence theorem} over a ball $B(r)$ of radius $r > 0$, centred at the origin in $\mathbb{R}^d$:
\begin{align*}
    \int (\mathcal{A}_P h) \mathrm{d}P 
    = \int \frac{1}{p} \nabla \cdot( p h ) \mathrm{d}P
    = \int \nabla \cdot ( p h ) \mathrm{d}V 
    = \lim_{r \rightarrow \infty} \oint_{B(r)} p h \cdot n \mathrm{d}\sigma.
\end{align*}
Here $\mathrm{d}V$ denotes the volume element in $B(r)$, $\mathrm{d}\sigma$ denotes the surface area element on the boundary of $B(r)$, and $n$ denotes the unit normal to the boundary of $B(r)$.
In order for this final term to vanish, it is sufficient that $\|p h \cdot n \|$ vanishes uniformly with respect to the surface area $\oint_{B(r)} \mathrm{d}\sigma$, which is $O(r^d)$.
Thus, if $h \colon \mathbb{R}^d \rightarrow \mathbb{R}^d$ and $\log p : \mathbb{R}^d \rightarrow \mathbb{R}$ are both continuously differentiable and the tail condition
\begin{equation}\label{eqn:Tail}
\| h(x) \| \leq C \| x \|^{-\delta} p(x)^{-1}
\end{equation}
is satisfied for some $C \in \mathbb{R}$, some $\delta > d -1$, and all $x \in \mathbb{R}^d$ outside of a bounded set, then the Stein identity is satisfied \citep{South2020}.
\end{textbox}

\begin{textbox}[t!]
\section{KERNEL STEIN DISCREPANCY}
To facilitate computation of the supremum in \Cref{eq: stein disc}, one can specialise to a particular form of Stein discrepancy called \textit{kernel Stein discrepancy}.
A \emph{kernel} is a symmetric, positive-definite function $k:\mathcal{X} \times  \mathcal{X} \rightarrow \mathbb{R}$.
A kernel $k$ \emph{reproduces} a Hilbert space, denoted $\mathcal{H}(k)$, whose inner product is denoted $\langle \cdot, \cdot \rangle_{\mathcal{H}(k)}$.
This means that the elements of $\mathcal{H}(k)$ are functions $f:\mathcal{X}\to\mathbb{R}$, and it holds that (i) $k(\cdot,x) \in \mathcal{H}(k)$ for all $x \in \mathcal{X}$, and (ii) $\langle f, k(\cdot,x) \rangle_{\mathcal{H}(k)} = f(x)$ for all $x \in \mathcal{X}$, $f \in \mathcal{H}(k)$. 
For example, the \textit{Gaussian} kernel $k(x,y) = \exp(-(x-y)^2)$ reproduces a Hilbert space that contains functions of the form $f(x) = \sum_{i=1}^m w_i \exp(-(x-y_i)^2)$ for all $w_i \in \mathbb{R}$, $y_i \in \mathbb{R}$, $m \in \mathbb{N}$, as well as certain limits of such functions \citep[][]{berlinet2011reproducing}.

The main observation here is that if we take the set $\mathcal{H} := \textstyle \{h:\mathbb{R}^d\rightarrow\mathbb{R}^d: \sum_{i=1}^d \langle h_i , h_i \rangle_{\mathcal{H}(k)} \leq 1 \}$, then the supremum in \Cref{eq: stein disc} can be exactly evaluated.
Let $\mathcal{H}(k)^d := \mathcal{H}(k) \times \cdots \times \mathcal{H}(k)$ denote the Cartesian product; i.e. the elements of $\mathcal{H}(k)^d$ are functions $h: \mathcal{X} \rightarrow \mathbb{R}^d$ with components $h_i \in \mathcal{H}(k)$.
Then \citet[][Theorem 1]{Oates2017} showed that the set of functions of the form $\mathcal{A}_P h$, $h \in \mathcal{H}(k)^d$ is a Hilbert space reproduced by the kernel 
\begin{equation}
k_P(x,y) := \nabla_x\cdot\nabla_y k(x,y) + \nabla_x k(x,y) \cdot u(y) + \nabla_y k(x,y) \cdot u(x) + k(x,y) u(x) \cdot u(y) , \label{eq: stein kernel}
\end{equation}
where $u(x) := \nabla \log p(x)$.
Assuming that, for each $\mathcal{A}_P h$, $h \in \mathcal{H}(k)^d$, the tail condition \Cref{eqn:Tail} is satisfied, then following \cite{liu2016kernelized} and \cite{chwialkowski2016kernel}, one can show that
\begin{equation}
    D\left( P , \sum_{i=1}^n w_i \delta(x_i) \right)^2 = \sum_{i=1}^n \sum_{j=1}^n w_i w_j k_P(x_i, x_j)  \label{eq: KSD def}
\end{equation}
for all $w_i \in \mathbb{R}$, $x_i \in \mathbb{R}^d$, $n \in \mathbb{N}$.
The kernel Stein discrepancy in \Cref{eq: KSD def} can therefore be exactly computed whenever the gradient $\nabla \log p$ can be evaluated.
Furthermore, under certain conditions, the kernel Stein discrepancy provides weak convergence control \citep[][Theorem 8]{gorham2017measuring}.

\end{textbox}

The optimisation problem we are interested in thus reduces to the problem of identifying weights $w_i$ and an index sequence $\pi$ for which the kernel Stein discrepancy $D(P,Q_M)$ is minimised when $Q_M$ is the discrete distribution in \Cref{eq: post processed}. 
Approaches based on Stein discrepancy minimisation include \emph{black-box importance sampling} \citep{liu2017black,Hodgkinson2020}, \emph{Stein points} \citep{Chen2018SteinPoints,chen2019stein}, and \emph{Stein thinning} \citep{riabiz2020optimal}. 
In what follows we describe the Stein thinning approach of \cite{riabiz2020optimal}, illustrated in Figure \ref{fig:MCMC_output_processed} in the setting with equal weights $w_i = \frac{1}{M}$, and defer to \citet{liu2017black,Hodgkinson2020,riabiz2020optimal} for discussion of the case in which weights are not equal.

\begin{marginnote}[]
\entry{Stein thinning}{An algorithm that selects representative states from MCMC output in order that a Stein discrepancy $D(P,Q_M)$ between the distributional target $P$ and the approximation $Q_M$ in \Cref{eq: post processed} is minimised.}
\end{marginnote}

Combinatorial optimisation to elicit an index sequence $\pi$ for which the kernel Stein discrepancy $D(P,Q_M)$ is minimised presents some technical challenges, which we defer discussion of until \Cref{subsubsec: nonmyop}.
Here we describe the simple, sequential approach called \textit{Stein thinning} that was explored in \cite{riabiz2020optimal}.
This involves constructing $\pi$ in a sequential, greedy manner, in which at iteration $1\leq j \leq M$, an index $\pi(j)$ is selected according to
\begin{equation}
    \pi(j) \in \underset{i \in \{1,\dots,N\}}{\mathrm{argmin}} D\left(P, \frac{1}{j} \left[ \delta(X_i) + \sum_{j'=1}^{j-1} \delta(X_{\pi(j')}) \right] \right) 
\end{equation}
or, equivalently, using the explicit form of kernel Stein discrepancy in \Cref{eq: KSD def} of the inset box,
\begin{equation}
    \pi(j) \in 
    \underset{i \in \{1,\dots,N\}}{\mathrm{argmin}} \quad \dfrac{k_P(X_i,X_i)}{2} + \displaystyle\sum\limits_{j'=1}^{j-1}k_P(X_{\pi(j')},X_i) . \label{eq: Stein thinning}
\end{equation}
This procedure has computational complexity $\mathcal{O}(NM^2)$, or possibly less (since it is possible for a state to be repeatedly selected and the relevant quantities to be cached). 

The main conceptual advantages of Stein thinning and related algorithms, compared to the standard post-processing techniques described in \Cref{ssec:burnin} and \Cref{ssec:thinning}, are that (a) they directly address the bias-variance trade-off, (b) they can correct for systematic bias in the MCMC output, (c) they can automatically identify and remove a burn-in period.
The main practical limitation of Stein thinning and related algorithms is that there are certain pathologies of Stein discrepancy, which occur when either (a) $P$ has distant high-probability regions, or (b) $P$ is high-dimensional (e.g. $d > 100$), either of which can lead to poor approximations when $M$ is small; see \cite{wenliang2020blindness}.
To illustrate the potential advantages of Stein thinning, we now present a special case of Theorem 3 in \cite{riabiz2020optimal}, which describes conditions under which the sequence generated using the Stein thinning algorithm in \Cref{eq: Stein thinning} produces a discrete approximation $Q_M$ that converges almost surely to $P$. 
Note in particular that the result does not assume that the Markov chain is $P$-invariant.

\begin{theorem*}[Bias correction for MCMC]
	Let $P$, $P'$ be probability distributions with positive and continuous densities $p$ and $p'$ on $\mathbb{R}^d$.
	Assume that the tails of $P$ are \emph{distantly dissipative} \citep[a relaxation of log concavity; see][]{gorham2019} and that $p$ is continuously differentiable on $\mathbb{R}^d$.
	Consider a $P'$-invariant, time-homogeneous Markov chain $(X_i)_{i \in \mathbb{N}}$ generated using a $V$-uniformly ergodic transition kernel, where $V(x) = \frac{p(x)}{p'(x)} \sqrt{d + \|\nabla \log p\|^2}$.
	Suppose that, for some $\gamma > 0$, the following moment condition is satisfied:
	\begin{align}
	\sup_{i \in \mathbb{N}} \textstyle \mathbb{E} \left[ \exp \left( \gamma \max\left( 1 ,  \frac{p(X_i)}{p'(X_i)} \right)^2 \left( d + \|\nabla \log p(X_i) \|^2 \right) \right) \right] & < \infty.  \label{eq: moment}
	\end{align}
	Let $\pi$ be an index sequence of length $m$ produced by \Cref{eq: Stein thinning} applied to the MCMC output $(X_n)_{n \leq N}$, where $k_P$ in \Cref{eq: stein kernel} is based on the \emph{inverse multi-quadric} kernel $k(x,y) = (1 + \|x-y\|^2)^{-1/2}$.
	If $M \leq N$ and the growth of $N$ is limited to at most $\log(N) = O(M^{\beta/2})$ for some $\beta < 1$, then, $\mathbb{P}$-almost surely, $Q_M = \frac{1}{M} \sum_{j=1}^M \delta(X_{\pi(j)})$ converges weakly to $P$ as $M,N \rightarrow \infty$.
\end{theorem*}

This result, and the related results in \cite{liu2017black,Hodgkinson2020}, weaken or remove the requirement to design Markov chains that are exactly $P$-invariant.
See also \cite{gramacy2010importance,radivojevic2020modified}.
Less formally, this result suggests that one may not need to run a Markov chain to convergence in order for its output to be useful.
On the other hand, the moment condition in \Cref{eq: moment} imposes a requirement that $P'$ cannot be too dissimilar to $P$ (informally, the Markov chain must explore the high density regions of $P$, albeit not necessarily with the same frequencies as would be expected if the chain was $P$-invariant).
Being a recent line of research, it remains to be seen whether these advances in the post-processing of MCMC output will in turn influence the design of algorithms for MCMC.
Software to perform Stein thinning, including packages for \texttt{R}, \texttt{Python} and \texttt{MATLAB}, is available at \url{stein-thinning.org}.

\subsubsection{Extensions to Stein Discrepancy Minimisation} \label{subsubsec: nonmyop}

\begin{figure}[t!]
   \centering
   \includegraphics[width=0.8\textwidth]{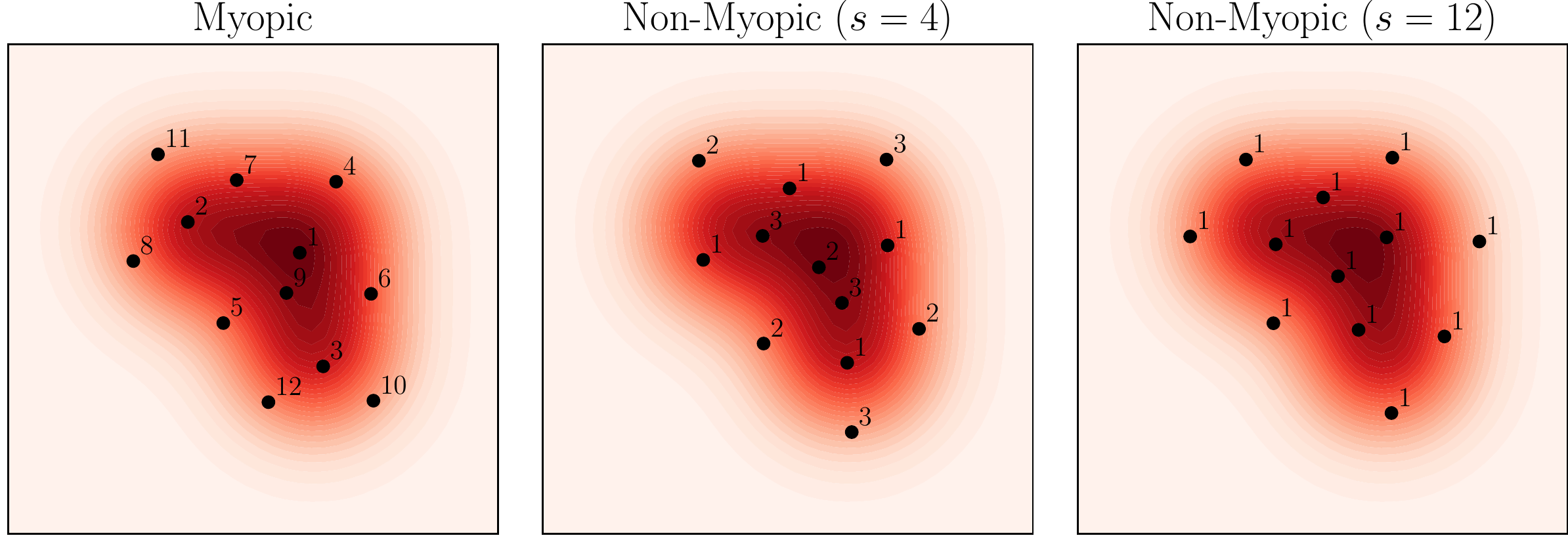}
   \caption{
   Extensions to Stein thinning:
    Here the sample path (gray line) in \Cref{fig:MCMC_output_processed} is post-processed to select $M=12$ representative states (black circles), to provide an approximation to the distributional target $P$, indicated by the shaded contour plot in the background.
    Left: Myopic selection of states (look ahead horizon $s=1$).
    Centre: Non-myopic selection of states, with a look-ahead horizon of $s=4$.
    Right: Non-myopic selection of states, with a look-ahead horizon of $s=12$. 
    [Integers indicate the iteration of the algorithm in which a given state was selected.]
   }
   \label{fig: stein thinning cartoon}
\end{figure}

The Stein thinning algorithm that we just described in \Cref{eq: Stein thinning} is \emph{myopic}, in that it selects the index of the single best state $\pi(j)$ at each iteration without consideration of whether this makes \emph{subsequent} choices better or worse overall. 
This myopia can make the optimisation statistically inefficient, as observed in the left panel of \Cref{fig: stein thinning cartoon}. Specifically, we see that the choice of the first state as the sample closest to the global mode of the distribution means that all possible choices for the second state will (temporarily) significantly worsen the overall approximation. 
In a less extreme fashion, this can also be seen for state 6, and state 8, by similar symmetry observations. 
A second shortcoming of the algorithm we just presented is that it requires scanning through the entire MCMC output of size $N$ at each iteration, which can lead to an unacceptable computational cost.

\begin{marginnote}[]
\entry{Non-myopic}{An optimisation algorithm is non-myopic if it looks further than a single step ahead when deciding the best course of action at a given iteration.}
\end{marginnote}

To ameliorate these shortcomings, at least to an extent, the Stein thinning algorithm can be generalised to both \emph{non-myopic} and \emph{mini-batch} settings as described in \cite{teymur2020optimal}.
The first of these extensions involves selecting multiple points simultaneously, while the mini-batch extension considers, at each iteration, selecting points from a random subset of the samples along the MCMC path. 
It was shown in \cite{teymur2020optimal} that these two extensions are \emph{synergistic}, in that non-myopic optimisation is most useful in the mini-batch context.
These extensions of Stein thinning will now be described.
Let $B \ll N$ be a \emph{mini-batch size} and let $(X_b^j)_{1\leq b \leq B,1 \leq j \leq M}$ be the collection of mini-batches, each of size $B$ and to be used at iteration $j$. 
For example, the mini-batches could be chosen uniformly, with or without replacement, from the MCMC output $(X_n)_{n \leq N}$. 
Let $s \in \mathbb{N}$ be the \emph{look-ahead horizon}, meaning the number of points to be simultaneously selected (the algorithm in \Cref{eq: Stein thinning} corresponds to $s=1$). 
Then we choose a vector $\pi(j,\cdot)$ of $s$ indices to be used at iteration $j$ by performing the optimisation
\begin{equation}
\pi(j,\cdot) \in \underset{S \in \{1,\dots,B\}^s }{\mathrm{argmin}} \bigg[ \displaystyle \frac{1}{2} \sum_{b,b' \in S} k_P(X_b^j,X_{b'}^j) 
 \displaystyle
+ \sum_{j'=1}^{j-1} \sum_{b=1}^s \sum_{b' \in S} k_P\big(X_{\pi(j',b)}^{j'},X_{b'}^j\big) \bigg] , \label{eq: nonmyop st}
\end{equation}
which we have obtained using the explicit form of kernel Stein discrepancy in the inset box, in a similar manner to how we obtained \Cref{eq: Stein thinning}. 
Run for $M$ iterations, this algorithm selects a representative set of $sM$ states, potentially with some states selected more than once. 
It is possible to apply a similar theoretical analysis to that in \Cref{ssec: stein disc min} to the generalised algorithm in \Cref{eq: nonmyop st}; see \cite{teymur2020optimal}.

\begin{marginnote}[]
\entry{Mini-batch}{A computationally advantageous approach in which the full set of candidate samples is parcelled up into small batches, and the selection algorithm is applied to each resulting subset separately.}
\end{marginnote}

\FloatBarrier

Implementation of this non-myopic algorithm requires solving the optimisation problem in \Cref{eq: nonmyop st}. 
This is a potentially challenging combinatorial optimisation problem, and is only tractable when the batch size $B$ is small (e.g. $B \leq 1000$). 
In order to solve it, one can represent the indices $S \subset \{1,\dots,B\}^s$ of the $s$ points to be selected at iteration $j$ as a vector $v^j \in \mathbb{N}_0^s := \{0,\dots,s\}^B$ whose $i$th element indicates the number of copies of $X_i^j$ that are selected from the $j$th mini-batch, where $v^j$ is constrained to satisfy $\sum_{i=1}^B v^j_i = s$. 
It is then an algebraic exercise to recast an optimal index sequence $\pi(j,\cdot)$ as the solution to a constrained \emph{integer quadratic programme} \citep[e.g.][]{wolsey21}:
\begin{equation}
\begin{gathered}[t]
 \underset{v^j \in \mathbb{N}_0^s}{\mathrm{argmin}} \ \dfrac{1}{2}v^{j\top} K_P^j v^j + c^{j\top} v^j \quad \text{such that} \quad \mathbf{1}^\top v^j = s,  \\[-0.2em]
[K_P^j]_{i,i'} := k_P(X^j_i,X^j_{i'}), \ \ \ \ 
c^j_i := \sum_{i'=1}^{i-1}\sum_{j'=1}^s k_P(X_{\pi(i',j')},X_j). 
\end{gathered}
\label{eq: iqp}
\end{equation}
\begin{marginnote}[]
\entry{Integer quadratic programme}{An optimisation problem in which the objective function is quadratic, and where the solutions are constrained to be integer-valued.}
\end{marginnote}

Depending on the values of $B$, $M$ and $N$, and the way in which the mini-batches are selected, it may be advantageous to store and reuse kernel calculations from iteration to iteration. In general, however, we assume that the matrix $K_P^j$ and vector $c^j$ are recalculated for each batch, giving the algorithm an overall complexity of $\mathcal{O}(M^2s^2B^s)$. This apparently daunting computational complexity can nevertheless be advantageous if $N$ is very large and $B \ll N$.
\cite{teymur2020optimal} recommends a ratio $s/B \approx 10$, though this is expected to be problem-dependent.
Finding the exact solution of this type of optimisation problem is NP-hard\footnote{Without the cardinality constraint $\mathbf{1}^\top v^j = s$, this integer quadratic programme is equivalent to the celebrated \emph{MAX-CUT} problem; and with this constraint to the related \emph{cardinality constrained $k$-partition} problem \citep{Rendl_2012}.}, however 
a `good' feasible solution may still be useful.
Indeed, the iterative nature of the overall algorithm allows it to compensate, to a degree, for sub-optimal selection of states at a given iteration through its selection of states in future iterations. 
Fortunately, `good' solutions can readily be obtained using any of a number of packaged discrete optimisation routines, such as 
the commercial software $\mathtt{gurobi}$, $\mathtt{MOSEK}$ and $\mathtt{MATLAB}$'s Optimization Toolbox, or numerous open-source equivalents.

\subsection{Summary}

This completes our review of post-processing strategies for MCMC, when the aim is to accurately approximate the distributional target $P$ itself.
Given that convergence diagnostics and thinning are well-known techniques, we deliberately focussed on their shortcomings in this review.
Then, we described recent methodology that aims to directly address the bias-variance trade-off that occurs when post-processing MCMC output.
This trade-off is fundamental to many important and challenging applications of MCMC, in which there is a practical limit to the computational budget.
To limit scope, we did not discuss alternative classes of algorithm, such as \textit{unbiased Monte Carlo} \citep{jacob2020unbiased}, for which a bias-variance trade-off is systematically avoided.
Finally, it was argued that recent developments in Stein discrepancy have the potential to substantially impact on both applications of, and research into, MCMC.

\section{APPROXIMATION OF POSTERIOR EXPECTATIONS} \label{sec: approx expect}

In contrast to exploratory Bayesian analyses, several applications of Bayesian statistics require just a finite number of scalar posterior quantities of interest.
For example, in a decision-making context, the Bayes rule may take an explicit and simple form, such as the mean of the posterior or perhaps a median, or a higher moment \citep{berger2013statistical}.
To proceed, one can first run MCMC, followed by suitable post-processing as described in \Cref{sec: approx posterior}, to obtain an approximation to the posterior from which quantities of interest can be extracted.
However, approximating the full posterior may incur unnecessary computational effort. 
In such circumstances it is natural to seek to focus computational resources on approximating just the quantities of interest.

\textit{Control variates} are a classical technique for reducing the variance of Monte Carlo estimators, which are used in a wide range of applications, including stochastic gradient-based optimisation \citep{wang2013variance,grathwohl2017backpropagation} and as part of MCMC methods themselves \citep{Baker2019}.
In this section we review the use of control variates as a post-processing technique for MCMC. 
It will be shown that modern control variates, unlike their classical counterparts, can facilitate bias removal, as well as variance reduction.
In \Cref{sec: CVs} the control variate technique is presented at a general level, then in \Cref{sec: gradient CVs} we present specific control variates techniques and explain how these can be used to post-process MCMC.

\subsection{Monte Carlo Estimators}
\label{sec: CVs}

For the purposes of this article, a \textit{Monte Carlo estimator} is a map $\mu : \Omega \times \mathcal{L}^2(P) \rightarrow \mathbb{R}$ 
whose output $\mu(\omega,f)$ depends on $\omega$ only via dependence on a collection of random variables $X_1(\omega),\dots,X_n(\omega)$. 
The output, $\mu(\omega,f)$, is interpreted as an approximation to the integral $\int f \mathrm{d}P$, which we consider to be a scalar quantity of interest.
Our focus is on Monte Carlo estimators that are based on MCMC output, with the standard example being the estimator
\begin{equation}
    \mu(\omega,f) = \frac{1}{N} \sum_{i=1}^N f(X_i(\omega)) , \label{eq: linear}
\end{equation}
that takes an average of $f$ over the states $(X_n)_{n \leq N}$, in the MCMC output.
Such an estimator is said to be \textit{consistent} if, for all $f \in \mathcal{L}^2(P)$, the random variable in \Cref{eq: linear} converges in probability to $\int f \mathrm{d}P$ as $N \rightarrow \infty$.
The asymptotics of \Cref{eq: linear} are well-studied in the setting where the Markov chain is $P$-invariant \citep{meyn2012markov}. 
Improved approximations can be obtained using the methods described in \Cref{sec: approx posterior}.
For example, post-processed MCMC output of the form in \Cref{eq: post processed}, can be used to provide a Monte Carlo estimator 
\begin{equation}
    \mu(\omega,f) = \sum_{i=1}^M w_i f(X_{\pi(i)}(\omega)) . \label{eq: improved MC}
\end{equation}
In the setting where the Markov chain is not $P$-invariant, \Cref{eq: linear} will be asymptotically biased in general but \Cref{eq: improved MC} may yet be consistent, as explained in \Cref{ssec:mindiscrepancy}, and therefore \Cref{eq: improved MC} may be preferred.
In the presence of several consistent estimators, it is natural to ask which estimator should be preferred; this question can be rigorously formulated in terms of the \textit{mean square error} of the estimators and the answer will be $f$-dependent in general.
For convenience we will leave the $\omega$ argument implicit in the remainder of this section.

\subsubsection{Selecting a Monte Carlo Estimator}

The \textit{mean square error} of a Monte Carlo estimator $\mu$ is defined as
\begin{equation}
    \text{MSE}(\mu(f)) := \mathbb{E}\left[ \left( \mu(f) - \int f \mathrm{d}P \right)^2 \right] . \label{eq: MSE}
\end{equation}
Presented with a collection $\{\mu^\theta\}_{\theta \in \Theta}$ of Monte Carlo estimators, say indexed by $\theta \in \Theta$, we would like to select an estimator for which $\text{MSE}(\mu^\theta(f))$ is minimised.
Let us assume that the mean square error can itself be consistently estimated based on the MCMC output, i.e. we have available an estimator $\widehat{\text{MSE}}(\mu(f) )$.
Then a general recipe to select a Monte Carlo estimator is as follows:

\begin{summary}[GENERAL RECIPE TO SELECT A MONTE CARLO ESTIMATOR]
\begin{enumerate}
\item Identify a collection of Monte Carlo estimators $\mu^\theta$, $\theta \in \Theta$.
\item For each estimator, compute $\widehat{\text{MSE}}( \mu^\theta(f))$.
\item Select $\hat{\theta}$ such that $\theta \mapsto \widehat{\text{MSE}}( \mu^\theta( f ) )$ is minimised.
\end{enumerate}
\end{summary}

There are at least three possible shortcomings with this general recipe, which will be discussed.
First, it is not clear how one should identify an appropriate set of Monte Carlo estimators; control variates provide an elegant solution that we discuss next in \Cref{subsubsec: cvs}.
Second, it may be a challenging to identify a suitable estimator for the mean squared $\widehat{\text{MSE}}$, since the underlying MCMC method may be complicated.
Options for this are discussed in \Cref{subsubsec: proxies}.
Third, estimation error in $\widehat{\text{MSE}}$ presents a challenge when there are many Monte Carlo estimators being compared, since with more estimators there is a greater chance of selecting a poor estimator due to bad luck.
A solution to this problem requires that the size of the set of candidate estimators is controlled in some way commensurate with the error in $\widehat{\text{MSE}}$.
Several solutions will be discussed in \Cref{sec: gradient CVs}, including restricting the size of this set through the use of explicit finite dimensional bases, and through coupling the size of $\Theta$ to the size $N$ of the MCMC output.

It is emphasised that, compared to the techniques reviewed in \Cref{sec: approx posterior}, the selection of Monte Carlo estimators remains as much an art as a science.
Theoretical analyses are available on some aspects of the general recipe just outlined, and will be highlighted, but to our knowledge there does not yet exist a theoretical treatment that is broadly applicable in the MCMC context.

\subsubsection{Constructing Monte Carlo Estimators Using Control Variates}
\label{subsubsec: cvs}

An element $g \in \mathcal{L}^2(P)$ is said to be a \textit{control variate} (for $P$) if $\int g \mathrm{d}P = 0$.
Clearly any finite linear combination of control variates is also a control variate, and we will use $\mathcal{G}$ to denote a linear subspace of $\mathcal{L}^2(P)$ whose elements are control variates.
The power of control variates is that they enable one to take a single Monte Carlo estimator, such as \Cref{eq: linear}, and from this to generate a possibly large collection of Monte Carlo estimators.
Indeed, armed with a consistent Monte Carlo estimator $\mu$ and a set of control variates $\mathcal{G}$, one can consider Monte Carlo estimators of the form $\mu^\theta(f) := \theta_1 + \mu(f - \theta_1 - \theta_2)$ where $\theta_1 \in \mathbb{R}$, $\theta_2 \in \mathcal{G}$, $\Theta = \mathbb{R} \times \mathcal{G}$.
The consistency of $\mu$ is automatically inherited by each $\mu^\theta$.

\begin{marginnote}[]
\entry{Control Variate}{A square-integrable function whose expectation is 0.}
\end{marginnote}

Up to this point we have not discussed how control variates can be found in practice. Many approaches for developing control variates in the context of Markov chain sampling are based on approximating the solution $\hat{f}$ to the typically intractable Poisson equation
\begin{align}\label{eqn:PoissonDiscrete}
\hat{f} - K\hat{f} = f - \mathbb{E}[f],
\end{align}
where $K$ is the one-step ahead prediction operator $K\hat{f} = E[\hat{f}(X^{(n+1)})|X^{(n)}=x]$. In this setting, one could evaluate $\mathbb{E}[f]$ exactly by evaluating $f + K\hat{f} - \hat{f}$. \citet{Andradottir1993} propose numerical algorithms to approximate this solution in the context of finite state spaces. \citet{henderson1997variance} approximates the solution for specific Markov samplers, focusing on continuous-time processes and applications in stochastic network theory. This was extended in \citet{dellaportas2012control} for reversible Markov chains where $K$ is tractable for some basis functions. A method to approximate the solution to the Poisson equation by discretising the state space for geometrically ergodic Metropolis--Hastings chains is introduced in \citet{mijatovic2018poisson}.

Control variates have also been built for independent Metropolis--Hastings samplers \citep{Atchade2005} and for general Metropolis--Hastings samplers \citep{hammer2008}, although the latter approach requires an extension of the state space to include proposals.

The aforementioned control variates are sampler-specific or require adjustments to the MCMC algorithm. \Cref{sec: gradient CVs} describes sampler-agnostic control variates that are applicable when $\nabla \log p$ or an unbiased estimate is available.

\subsubsection{Proxies for Mean Square Error}
\label{subsubsec: proxies}

The problem of estimating the mean square error of a Monte Carlo estimator is difficult, due to the fact that both the dependence between the states $(X_n)_{n \leq N}$ in MCMC output, and the way that these states are combined in the Monte Carlo estimator $\mu$, can be arbitrarily complicated. 
See, for example, \citet{Flegal2010} for strategies that can be used to estimate the mean square error of the Monte Carlo estimator in \Cref{eq: linear}. 
To promote generality, here we consider simple and generic proxies for mean square error that are easily computed, 
and much of what we recommend is based on empirical evidence only. 

A simple proxy for mean square error can be obtained by considering \Cref{eq: improved MC} in the idealised setting where $X_i \sim P$, for which it follows
\begin{equation}
\text{MSE}\left( \sum_{i=1}^M w_i f(X_{\pi(i)})  \right) \leq \frac{1}{M} \int \left( f - \int f \mathrm{d}P \right)^2 \mathrm{d}P =: \frac{ \text{Var}(f) }{M}  , \label{eq: var}
\end{equation}
with equality when the $X_i$ are independent.
The variance $\text{Var}(f)$ can be estimated using the \textit{empirical variance}  
\begin{equation}
\widehat{\text{Var}}(f) := \sum_{i=1}^M w_i \left( f(X_{\pi(i)}) - \sum_{j=1}^M w_j f(X_{\pi(j)}) \right)^2 , \label{eq: emp var}
\end{equation}
evaluated using MCMC output.
Empirical variance minimisation for constructing control variates was studied in \cite{belomestny2017variance} for the case where the $X_i$ are independent.
For non-independent $X_i$, arising as MCMC output, a more involved proxy based on spectral approximation of the \textit{asymptotic variance} was studied in \cite{Brosse2019,belomestny2020varianceb,belomestny2020variancea}, representing probably the most successful attempt to-date to provide theory for control variates for post-processing MCMC output.
On the other hand, a popular and simple upper bound on \Cref{eq: emp var} is the \textit{least squares} estimator
\begin{equation}
    \widehat{\text{LS}}(f) :=  \sum_{i=1}^M w_i f(X_{\pi(i)})^2. \label{eq: LS}
\end{equation}
An empirical comparison of empirical variance and least squares estimators for the selection of control variates in  \cite{Si2020} reported that, perhaps surprisingly, the least squares estimator performed best.
That is, one selects $\hat{\theta} \in \mathbb{R} \times \mathcal{G}$ in order that $\theta \mapsto \widehat{\text{LS}}(f - \theta_1 - \theta_2)$ is minimised.
The scalar integral of interest is then estimated as
\begin{equation}
    \int f \mathrm{d}P \approx \mu^{\hat{\theta}}(f) = \hat{\theta}_1 + \underbrace{ \sum_{i=1}^M w_i \left( f(X_{\pi(i)}) - \hat{\theta}_1 - \hat{\theta}_2(X_{\pi(i)}) \right)  }_{=0}, 
    \label{eq: fit cv}
\end{equation}
where we have used the defining optimality property of $\hat{\theta}_2$ to conclude that the summation in \Cref{eq: fit cv} is zero.
One can equivalently describe this estimator as the result of first solving the \textit{weighted least squares} regression problem
\begin{equation}
    f(x_i) = \theta_1 + \theta_2(x_i) + \epsilon_i \label{eq: cv as regression}
\end{equation}
for the intercept $\theta_1 \in \mathbb{R}$ and the predictor $\theta_2 \in \mathcal{G}$, where the dataset consists of the (random) covariates $x_i = X_{\pi(i)}$ and independent errors $\epsilon_i \sim \mathcal{N}(0,w_i^{-1})$, $i = 1,\dots,M$, then reporting the fitted intercept $\hat{\theta}_1$ as an approximation to the integral of interest.
Next we address the question of how a set $\mathcal{G}$ of control variates can actually be constructed.

\subsection{Gradient-Based Control Variates}
\label{sec: gradient CVs}
Perhaps the main challenge in the application of control variate is identifying a suitable linear subspace $\mathcal{G}$. 
The elements of $\mathcal{G}$ should 
(i) have known expectation under $P$,
(ii) be easy to compute, and
(iii) offer an improvement on a Monte Carlo estimator $\mu$ that would otherwise have been used, in the sense that $\text{MSE}(\mu^{\theta}(f))<\text{MSE}(\mu(f))$ for some $\theta \in \mathbb{R} \times \mathcal{G}$. 
In this section we discuss gradient-based control variates that often meet these requirements, focussing on domains $\mathcal{X} = \mathbb{R}^d$ for some $d \in \mathbb{N}$.
These gradient-based control variates are constructed using mathematical tools similar to those exploited in \Cref{sec: approx posterior}.
The construction of control variates for non-Euclidean domains is discussed in \cite{Barp2018} for closed manifolds, while the general case, including discrete domains, remains under-developed.

Recall the operator $\mathcal{A}_P$ defined in \Cref{eq: Ap}; i.e. $\mathcal{A}_P h = \nabla \cdot h + \nabla \log p \cdot h$ where $h : \mathbb{R}^d \rightarrow \mathbb{R}^d$.
It was shown (in the inset box) that $\int \mathcal{A}_P h \; \mathrm{d}P = 0$ under an appropriate tail condition on $h$; it therefore is natural to consider a linear subspace of $\mathcal{L}^2(P)$ consisting of control variates of the form $\mathcal{G} = \mathcal{A}_P \Phi = \{\mathcal{A}_P \phi: \phi \in \Phi\}$ 
where $\Phi := \{\phi: \mathbb{R}^d \rightarrow \mathbb{R}^d\}$ is a linear space of functions for which the aforementioned tail condition is satisfied. 
The form for gradient-based control variates described here can be traced to the physics literature \citep{Assaraf1999,Assaraf2003}, and was first brought to bear on MCMC in \citet{Mira2013}. 
As discussed in \Cref{sec: approx posterior}, the required gradients are produced as a by-product when gradient-based samplers, such as the Metropolis-adjusted Langevin algorithm \citep{Roberts2002} or Hamiltonian Monte Carlo \citep{Duane1987}, are used, making the combination of gradient-based sampling and gradient-based control variates particularly appealing \citep{Papamarkou2014}.

It remains to discuss how the set $\Phi$ of differentiable vector fields $\phi : \mathbb{R}^d \rightarrow \mathbb{R}^d$ can be selected.
In what follows we review some of the main choices that previous researchers have considered.

\subsubsection{Finite-Dimensional Basis}\label{ssec:ZVCV}

Perhaps the simplest choice for $\Phi$ is the linear span of a finite set $\{\mathbf{\phi}_1,\ldots,\mathbf{\phi}_{J}\}$. 
There is clearly much flexibility in the choice of the vector fields $\phi_j$, but a popular choice is to use the gradients of monomials. 
Specifically, the so-called \textit{zero-variance control variates} (ZVCV) of \citet{Assaraf1999,Assaraf2003,Mira2013} sets $\Phi$ to be gradients of the class of $r$-th order polynomials, $\Phi=
\text{span}\{\nabla \mathbf{x}^{\alpha}:\alpha \in \mathbb{N}_0^d, 0 < |\alpha|\leq r\}$ where $r\in \mathbb{N}$, $\mathbf{x}^{\mathbf{\alpha}} = \prod_{i=1}^d x_i^{\alpha_i}$ and $|\mathbf{\alpha}|=\sum_{i=1}^d |\alpha_i|$. 
The number of basis functions is therefore $J={d+r \choose d}$ 
and the associated set $\mathcal{G}$ of control variates contains elements of the form
\begin{equation}\label{eqn:ZVCV_formula}
   \mathcal{A}_P (\nabla \mathbf{x}^{\alpha}) =  \sum_{j=1}^d \alpha_j\left[ (\alpha_j -1) x_j^{\alpha_j-2} + x_j^{\alpha_j-1}\nabla_{x_j}\log p(x) \right] \prod_{i \neq j} x_i^{\alpha_i}.
\end{equation}

Having identified $\Phi$, we can aim to select an optimal control variate from $\mathcal{G}$ using one of the proxies for mean square error discussed in \Cref{subsubsec: proxies}. 
Suppose that $J < M$ and consider the least squares proxy in \Cref{eq: LS}.
In what follows we consider a Monte Carlo estimator of the form in \Cref{eq: improved MC}, which of course contains, as a special case, the \textit{vanilla} Monte Carlo estimator in \Cref{eq: linear}.
Then we solve the regression problem in \Cref{eq: cv as regression} to obtain a fitted regression model
\begin{align}\label{eqn:regression}
\hat{f}(x) 
&= \hat{\theta}_1 + \sum_{j=1}^{J} \hat{\theta}_{2,j} \mathcal{A}_P \mathbf{\phi}_j(x),
\end{align}
where we collect the regression coefficients together into a vector $\mathbf{c} = (\hat{\theta}_1,\hat{\theta}_{2,1},\ldots,\hat{\theta}_{2,J})^\top \in \mathbb{R}^{J+1}$.
For completeness we now provide an explicit formula for the coefficient vector $\mathbf{c}$ in the fitted model. 
Let 
\begin{equation*}
\mathbf{f} = \begin{bmatrix} f(X_{\pi(1)}) \\ \vdots \\ f(X_{\pi(M)}) \end{bmatrix} , \; \;
W = \begin{bmatrix} w_1 & & \\ & \ddots & \\ & & w_M \end{bmatrix}, \; \;
\Phi = \begin{bmatrix} 1 & \mathcal{A}_P \mathbf{\phi}_1(X_{\pi(1)}) & \cdots & \mathcal{A}_P \mathbf{\phi}_{J}(X_{\pi(1)}) \\ \vdots & \vdots & \ddots & \vdots \\ 1 & \mathcal{A}_P \mathbf{\phi}_1(X_{\pi(M)}) & \cdots & \mathcal{A}_P \mathbf{\phi}_{J}(X_{\pi(M)}) \end{bmatrix}.
\end{equation*}
Then standard calculations show that selecting $\mathbf{c}$ to minimise $\widehat{\text{LS}}(f - \hat{f})$ leads to the estimated coefficients being $\hat{\mathbf{c}} = (\Phi^{\top} W \Phi)^{-1}\Phi^{\top} W \mathbf{f}$.
The integral $\int f \mathrm{d}P$ of interest is approximated by $\hat{\theta}_1$, the first component of $\mathbf{c}$.
The Monte Carlo estimator so obtained will be denoted $\mu^{\text{ZVCV}}(f) = \hat{\mathbf{c}}^\top \mathbf{e}_1 = \hat{\theta}_1$ in the sequel.

A intriguing property of gradient-based control variates with finite-dimensional bases is that, under many of the proxies for mean square error that we discussed, 
the resulting Monte Carlo estimators are \textit{semi-exact}, in the sense that $\text{MSE}(\mu^{\text{ZVCV}}(f))=0$ when $f \in \text{span} \{1\} \oplus \mathcal{A}_P \Phi$. 
\begin{marginnote}[]
\entry{Semi-exact}{A Monte Carlo estimator is semi-exact if it is exact on a linear subspace of $\mathcal{L}^2(P)$.}
\end{marginnote}
Recalling that for a Gaussian $P$, the gradient $\nabla \log p$ is a first order polynomial, semi-exactness in this case carries the interpretation of being exact for polynomials up to a certain order when $\Phi$ consists of gradients of monomials, in a similar way to how Gaussian cubature methods are constructed. 
This explains the ``zero variance'' nomenclature used in \citet{Assaraf1999,Assaraf2003,Mira2013}.

The main problem with using a finite-dimensional basis is that the regression problem is typically misspecified, since $f \notin \text{span} \{1\} \oplus \mathcal{A}_P \Phi$ for most functions $f$ of interest. 
This limits the variance reduction that can be achieved. 
To improve convergence rates, one could consider increasing the size of $\Phi$ with increasing $M$, in the spirit of \citet{portier2018monte,South2018}, or using an infinite-dimensional basis with regularisation, as described next.

\subsubsection{Infinite-Dimensional Basis}\label{ssec:CF}

\citet{Oates2017} extended the gradient-based control variates of \citet{Assaraf1999,Assaraf2003,Mira2013} to an infinite-dimensional linear subspace of $\mathcal{L}^2(P)$. 
This was achieved by taking $\Phi =  \mathcal{H}(k)^d$ to be a Cartesian product of reproducing kernel Hilbert spaces $\mathcal{H}(k)$ of sufficiently regular functions; see the inset box in \Cref{sec: approx posterior} for background.
The resulting set of control variates is $\mathcal{G}=\mathcal{A}_P \Phi=\mathcal{H}(k_P)$, which is again a reproducing kernel Hilbert space with reproducing kernel $k_P(x,y)$ defined in \Cref{eq: stein kernel}. 
The resulting method was referred to as \textit{control functionals} (CF), being a non-parametric (or `functional') generalisation of existing control variates.

The major challenge associated with an infinite-dimensional set $\mathcal{G}$ of control variates is \textit{over-fitting}; there may be infinitely many $\theta \in \mathbb{R} \times \mathcal{G}$ for which $\widehat{\text{MSE}}(\mu^\theta(f)) = 0$, yet in reality $\text{MSE}(\mu^\theta(f))$ may be arbitrarily large.
Consider, for instance, the least squares proxy 
\begin{align}
\widehat{\text{LS}}(f - \theta_1 - \theta_2) = \sum_{i=1}^M w_i \left( f(X_{\pi(i)}) - \theta_1 - \theta_2(X_{\pi(i)}) \right)^2,  \label{eq: LS interp}
\end{align}
which can be driven to zero by taking $\theta_2$ to interpolate $f - \theta_1$ at the nodes $X_{\pi(i)}$, $i = 1,\dots,M$.
Constraining $\theta_2$ at a finite set of locations does not constrain what $\theta_2$ may do outside this finite set, and is therefore not sufficient to provide control on $\text{MSE}(\mu^\theta(f))$.
The methodological contribution of \citet{Oates2017} was to select, among the set of $\theta \in \mathbb{R} \times \mathcal{G}$ for which \Cref{eq: LS interp} is minimised, an element with minimal semi-norm, where the semi-norm on $\mathbb{R} \times \mathcal{G}$ was defined as $|\theta|^2 = \langle \theta_2 , \theta_2 \rangle_{\mathcal{H}(k_P)}$. 
Under regularity assumptions, it can be shown that there exists a unique such element $\theta \in \mathbb{R} \times \mathcal{G}$.
Moreover, there is a closed-form solution to this optimisation problem which leads to the estimator $\mu^{\text{CF}}(f) = (\mathbf{1}^{\top} K_P^{-1} \mathbf{1})^{-1}(\mathbf{1}^{\top} K_P^{-1} \mathbf{f})$, where  $[K_P]_{i,j} = k_P(X_{\pi(i)},X_{\pi(j)})$. 
Note that we may without loss of generality assume that the $X_{\pi(i)}$ are distinct in \Cref{eq: improved MC}, since otherwise we could consider smaller $M$ and modify the weights accordingly.
This ensures that the matrix $K_P$ is non-singular whenever $k_P$ is a genuine reproducing kernel.
An interesting feature, and possible weakness, of CF is that the Monte Carlo estimator obtained does not depend on the weights $w_i$ appearing in \Cref{eq: improved MC}.

The performance of CF is heavily dependent on the choice of the kernel $k$. 
A common choice is to use a radial kernel $k$, such that $k(x,y)$ depends only on $\|x-y\|$, with examples including the Gaussian, Mat\'ern and rational quadratic kernels \citep{rasmussen2003gaussian}. 
Typically such kernels will be parametric, with a small number of parameters $\ell$ that must be specified.
\citet{Oates2017} recommended using cross-validation to select kernel parameters $\ell$, wherein a subset of the $\{ X_{\pi(i)}$, $i \in I_{\text{train}} \}$, are used construct the Monte Carlo estimator $\mu^{\theta}(f)$ where $\theta = \hat{\theta}^\ell \in \mathbb{R} \times \mathcal{G}$ and performance of this Monte Carlo estimator associated with $\ell$ is measured by the sum of squared errors $E_\ell := \sum_{i \in I_{\text{test}}} w_i (f(X_{\pi(i)}) - \hat{\theta}_1^\ell - \hat{\theta}_2^\ell(X_{\pi(i)}))^2$, where $I_{\text{test}} = \{1,\dots,M\} \setminus I_{\text{train}}$.
One then selects the kernel parameters $\ell$ for which $E_\ell$ is minimised.

Under regularity assumptions, CF has theoretical advantages over ZVCV. \citet{Oates2019,Barp2018} used results from scattered data approximation \citep{wendland2004scattered} to prove that, in the uniformly weighted case (i.e. $w_i = \frac{1}{M}$), the expected error $\mathbb{E}[ | \mu^{\hat{\theta}}(f) - \int f \mathrm{d}P | ]$ converges at a rate $O(M^{-s/d} \log(M)^{-s/d})$, where here $s$ is the number of (weak) derivatives of the function $f$ whose integral is sought. 
This indicates that the use of CF for post-processing MCMC output can actually improve the convergence rate of the estimator compared to standard MCMC, provided that the smoothness $s$ of $f$ is commensurate with the dimension $d$ of the domain on which it is defined (i.e. $s > \frac{d}{2}$). 
CF is thus an example of a method that offers \textit{super-$\sqrt{M}$ convergence}. 
\begin{marginnote}[]
\entry{Super-$\sqrt{M}$ convergence}{The property of having a convergence rate that is $o(M^{-1/2})$.}
\end{marginnote}
The main weakness of CF is that its performance can be inferior to ZVCV when the dimension $d$ is high relative to the size $M$ of the dataset; next we discuss how this weakness can be addressed.

\subsubsection{Mixed Basis}\label{ssec:SECF}

To address the poor performance of CF relative to CV in the high-dimensional context, \citet{South2020} generalised the approaches discussed in \Cref{ssec:ZVCV} and \Cref{ssec:CF}, to consider functional approximations of the form
\begin{align}
\hat{f}(x) = \hat{\theta}_1 + \hat{\theta}_2(x) + \sum_{j=1}^{J} \hat{\theta}_{2,j}' \mathcal{A}_P \phi_j (x), 
\label{eq: mixed basis}
\end{align}
where the parameters $\hat{\theta}$, consisting of $\hat{\theta}_1 \in \mathbb{R}$, $\hat{\theta}_2 \in \mathcal{H}(k_P)$ and $\hat{\theta}_2' \in \mathbb{R}^{J}$, are about to be specified.
Notice that one recovers the same form of approximation used in ZVCV, i.e. \Cref{eqn:regression}, as the special case where $\hat{\theta}_2 = 0$.
Similarly one can show that the same form as CF is recovered when $\hat{\theta}_2'=0$. 
Thus \Cref{eq: mixed basis} represents a strict generalisation of ZVCV and CF, and one may hope to obtain the `best of both worlds', in terms of the superior performance of ZVCV in high dimensions and the super-$\sqrt{M}$ convergence of CF.
The performance of this hybrid approach depends on how the parameters $\hat{\theta}$ are selected. Following \citet{Sard1949}, \citet{South2020} propose to select $\hat{\theta}$ such that the following properties are satisfied:
\begin{enumerate}
\item $\hat{f} = f$ for all $f \in \text{span}\{1\} \oplus \mathcal{A}_P \Phi $, where $\Phi = \text{span}\{\phi_1,\dots,\phi_{J}\}$ \label{item:exactness}
\item $\widehat{\text{LS}}(f - \hat{f}) = 0$
\item $\hat{\theta}_2$ minimises $\theta_2 \mapsto \langle \theta_2 , \theta_2 \rangle_{\mathcal{H}(k_P)}$ subject to the first two properties being satisfied.
\end{enumerate} 
The first property is to ensure semi-exactness and the second is an interpolation requirement. 
The third property amounts to minimising the semi-norm $|\theta| = \langle \theta_2 , \theta_2 \rangle_{\mathcal{H}(k_P)}$ and serves to ensure uniqueness of $\hat{\theta}$ and to penalise complexity, similarly to CF.
This method is referred to as a \textit{semi-exact control functional} (SECF) and the closed-form solution for the estimator is $\mu^{\text{SECF}}(f) = \mathbf{e}_1^\top (\Phi^{\top}K_P^{-1}\Phi)^{-1} \Phi^{\top} K_P^{-1} \mathbf{f} $. 
If there are parameters $\ell$ in the kernel $k$ that must be specified, then cross validation can be applied in a similar way to that described in \Cref{ssec:CF}.
Similarly to CF, a possible weakness of this hybrid approach is that the Monte Carlo estimator obtained does not depend on the weights $w_i$ appearing in \Cref{eq: improved MC}.

\citet{South2020} demonstrated that such a hybrid approach can indeed enjoy the advantages of both ZVCV and CF; we illustrate this below in \Cref{para: choosing}.
Open-source software is available for ZVCV, CF and SECF, via the \texttt{ZVCV} package \citep{rZVCV} on the comprehensive R archive network (CRAN). The required input for this package is a set of $M$ samples and the associated evaluations of $f(\cdot)$ and $\nabla \log p(\cdot)$.

\subsubsection{Practical Considerations}\label{ssec:PracticalCV}

Earlier we alluded to the construction of control variates being more an `art' than a science; here we provide practical recommendations based on our personal experience using control variates to post-process MCMC.

\paragraph{Choosing a Control Variate Method}
\label{para: choosing}

Choosing between various control variate methods, like ZVCV, CF and SECF, is non-trivial. 
Cross-validation approaches are computationally expensive and prone to incorrect decisions due the need to reduce the sample size in each fold. It would therefore be helpful to have an understanding of the theoretical properties of different methods. Unfortunately, such theoretical analyses are under-developed at present. 
Specifically, the theory that does exist tends to involve assumptions that are difficult to verify in practice, if they hold at all. 
Table \ref{tab:CVcomparison} summarises the current state of knowledge for the methods that we have discussed. 

\begin{marginnote}[]
\entry{Bias-correcting}{Capable of removing asymptotic bias in certain biased MCMC algorithms.}
\end{marginnote}

\begin{table}[h]
\tabcolsep7.5pt
\caption{Properties of the control variate methods we have discussed.} 
\label{tab:CVcomparison}
\begin{center}
\begin{tabular}{|l|c|c|c|c|}
\cline{2-5}
 \multicolumn{1}{c}{} & \multicolumn{1}{|c|}{Complexity}    & Semi-exact  & Bias-correcting & Super-$\sqrt{M}$ \\
\hline
Vanilla MCMC & $O(Nd)$  & No  & No   & No \\
ZVCV + $\widehat{\text{LS}}$ & $O(Nd+Md^{2r} + d^{3r})$  & Yes  & No   & No\\
CF & $O(Nd+M^3+M^2d)$ & No & Yes  & Yes \citep[e.g.][]{Barp2018}\\
SECF & $O(Nd+M^3 + d^{3r})$   & Yes  & Yes & Yes (conjectured) \\
\hline
\end{tabular}
\end{center}
\end{table}

The positive entries in \Cref{tab:CVcomparison} should be interpreted as there being (possibly strong) theoretical assumptions under which the result has been established.
The fact that CF and SECF are \textit{bias-correcting} will not come as a surprise to the reader in light of the discussion in \Cref{ssec: stein disc min}.
A perhaps more useful approach to selection of a control variate method is to explore their empirical performance in the context of a synthetic test-bed.

\begin{figure}[t!]
\centering
\begin{subfigure}[t]{0.35\textwidth}
\includegraphics[width=1\textwidth]{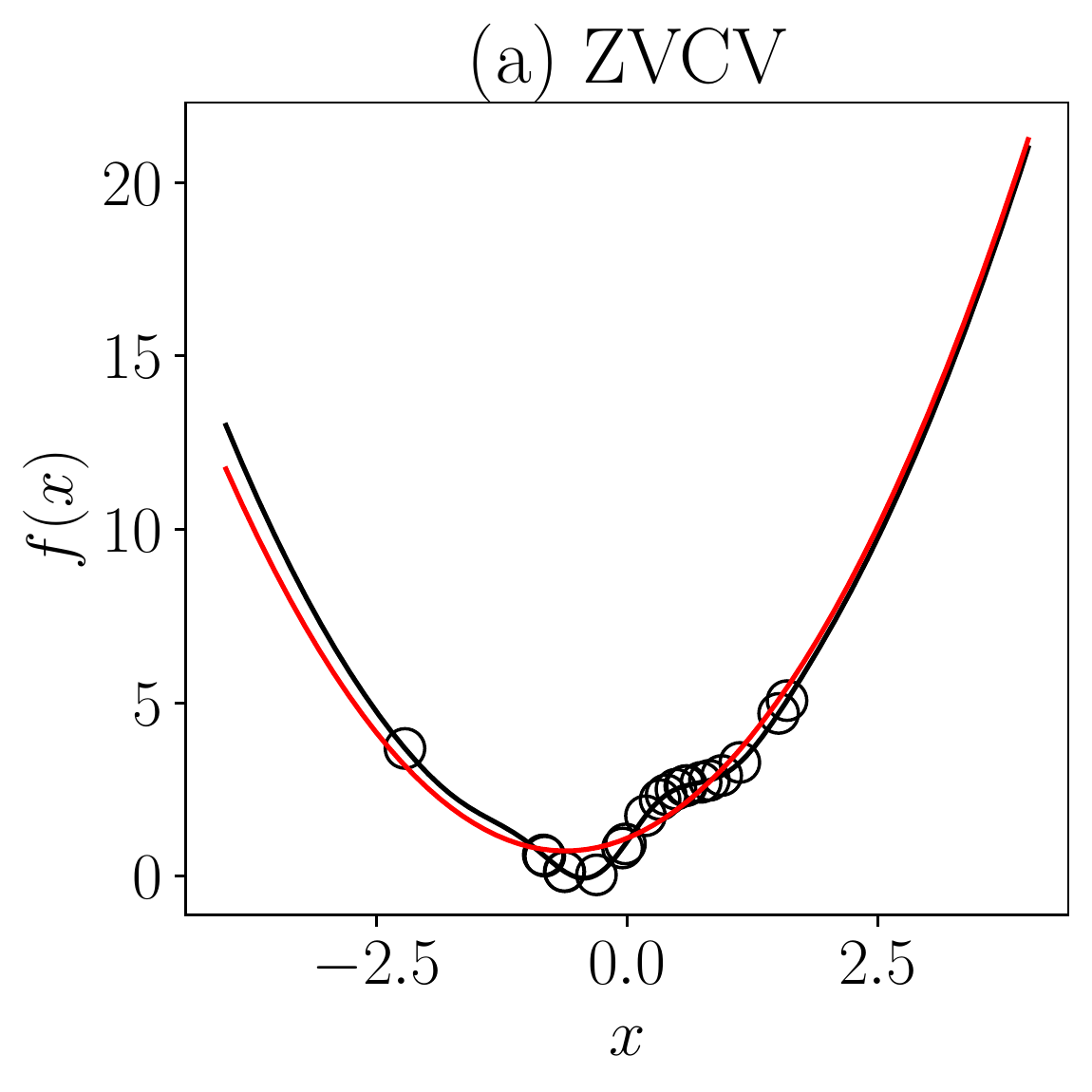}
\label{subfig:gauss_zvcv}
\end{subfigure}
\begin{subfigure}[t]{0.35\textwidth}
\includegraphics[width=1\textwidth]{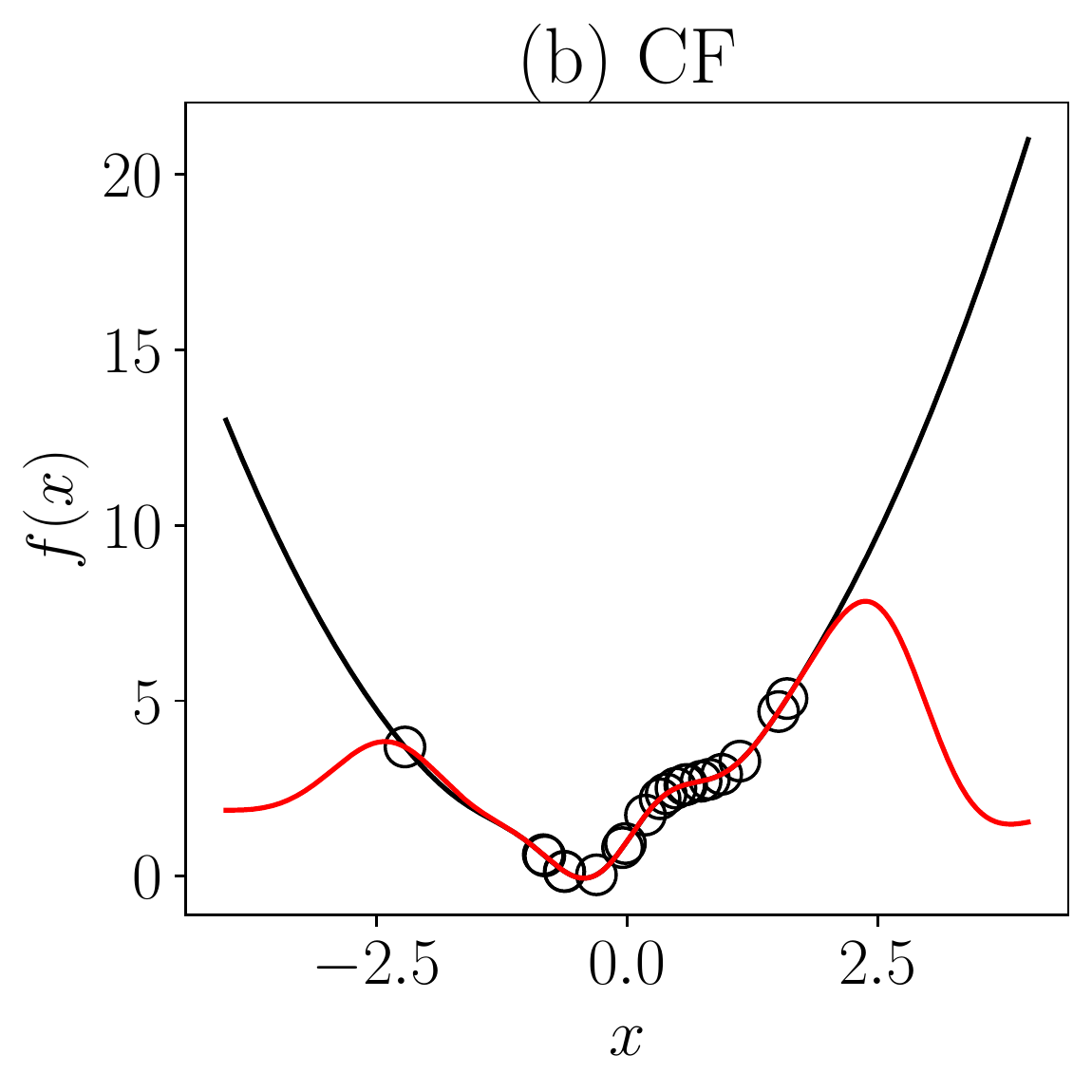}
\label{subfig:gauss_cf}
\end{subfigure}\\
\begin{subfigure}[t]{0.35\textwidth}
\includegraphics[width=1\textwidth]{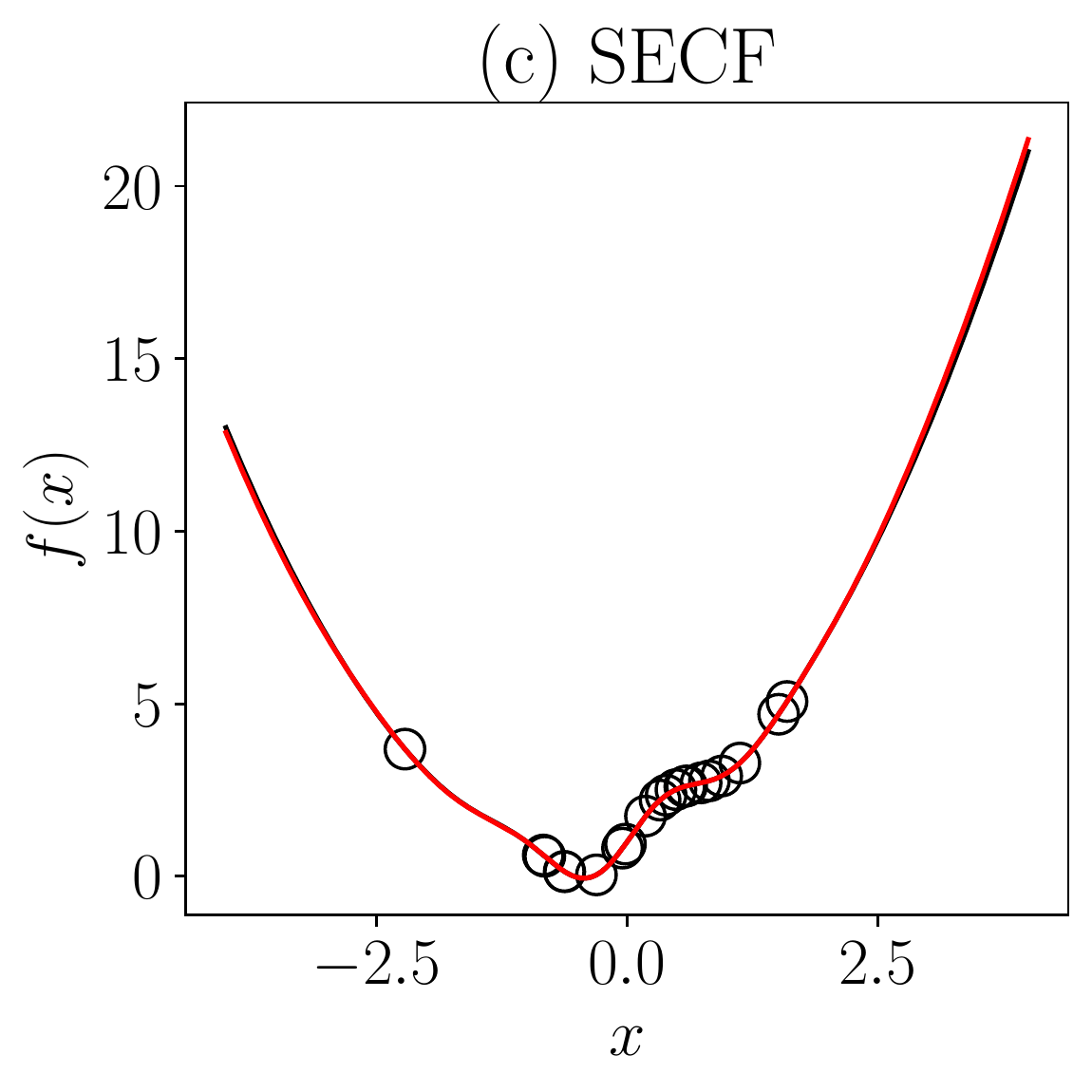}
\label{subfig:gauss_secf}
\end{subfigure}
\begin{subfigure}[t]{0.35\textwidth}
\includegraphics[width=1\textwidth]{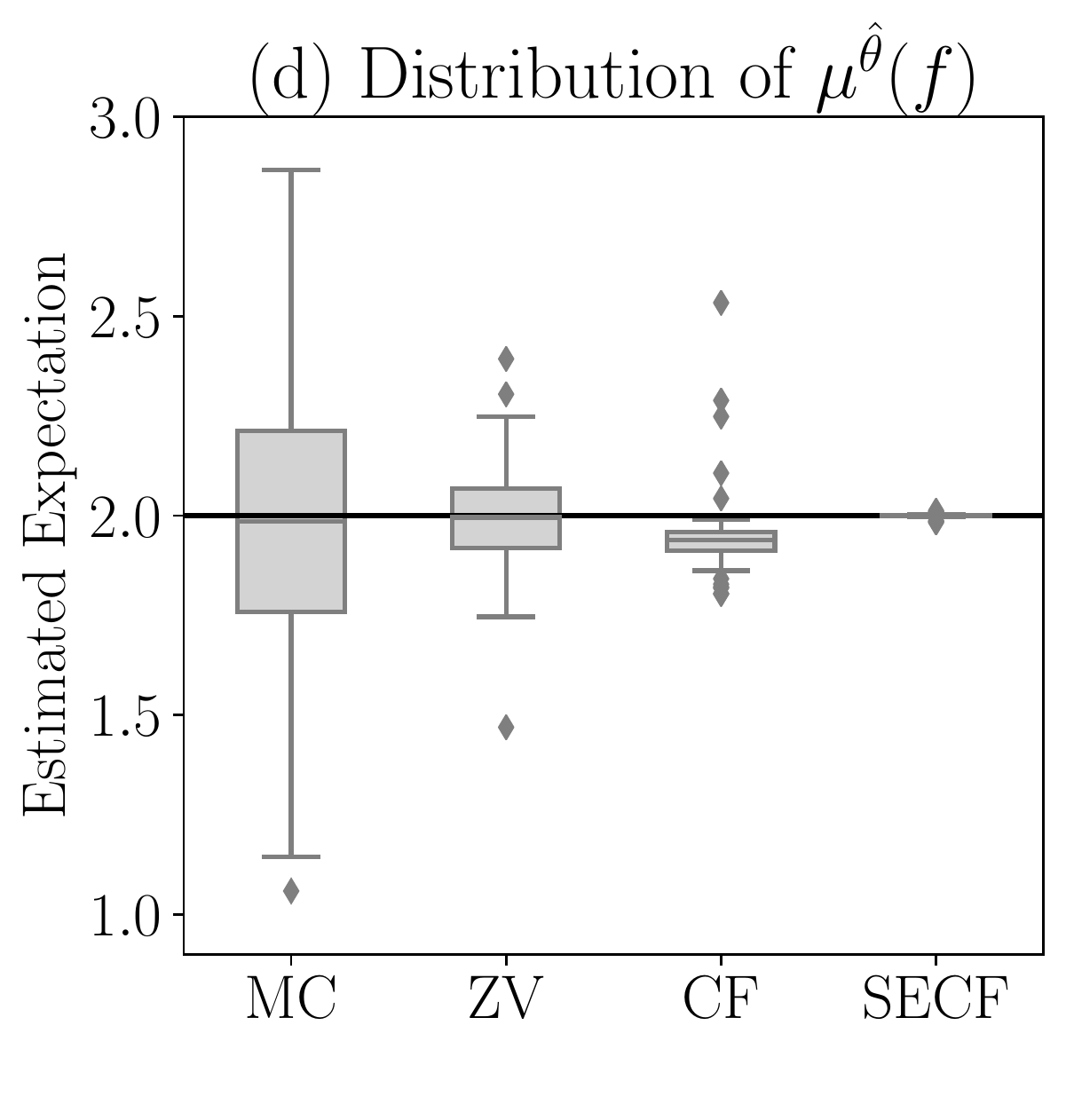}
\label{subfig:gauss_boxplot}
\end{subfigure}
\caption{Gradient-based control variates in a toy example. 
Figures 5a to 5c
show the function $f$ of interest (black line) along with the values $f(X_i)$ computed at the random locations $(X_i)_{i \leq N}$, $N=20$. 
These data are used to construct approximations $\hat{f}$ (red line) to $f$, in each of the methods ZVCV, CF and SECF. 
\Cref{subfig:gauss_boxplot} shows boxplots of 100 independent estimates for the integral $\int f \mathrm{d}P$ of interest.
}
\label{fig:GradientCV}
\end{figure}

\begin{example*}
Here we compare the performance of different control variate methods on a simple toy example that aims to represent the (relatively common) situation in which $P$ is approximately Gaussian, which may hold in applications for which there is a Bernstein-von-Mises limit. 
For illustrative purposes, we use a $1$-dimensional unit Gaussian distribution with density $p(x) = (2\pi)^{-1/2} \exp(-x^2/2)$ and we estimate the posterior expectation of $f(x) = 1 + x + x^2 + \sin(\pi x)\exp(-x^2)$, for which one can verify $\int f \mathrm{d}P = 2$. 
This function $f$ was chosen because the combination of complex behaviour near $x=0$ and polynomial behaviour in the tails present challenges for both the parametric and non-parametric methods.
For simplicity we consider an idealised MCMC algorithm that samples $X_i$ independently from $P$, and we consider the vanilla Monte Carlo (MC) estimator $\frac{1}{N} \sum_{i=1}^N f(X_i)$ as our starting point; i.e. we seek to reduce the variance of this Monte Carlo estimator using a control variate method.

The results are shown in \Cref{fig:GradientCV}. 
Here the approximating function $\hat{f}$ for ZVCV is 
a second order order polynomial\footnote{
ZVCV was implemented with a polynomial basis of order $r=2$, fit using $\widehat{\text{LS}}$. The form of $\hat{f}$ as a second order polynomial can be derived using \Cref{eqn:ZVCV_formula} and $\nabla \log p = -x$.}, which provides a poor approximation in the region where there are data but provides a good approximation in the tail (\Cref{subfig:gauss_zvcv}). 
For CF\footnote{CF was implemented with a Gaussian kernel $k(x,y) = \exp(-x^2/\lambda^2)$ where $\lambda$ is selected from $10^{\{-2,-1,0,1,2\}}$ using $3$-fold cross-validation.}, the interpolant $\hat{f}$ performs well in regions where there are data, less so in the tail (\Cref{subfig:gauss_cf}). 
In contract, SECF\footnote{SECF was implemented with $r=2$ and a Gaussian kernel $k(x,y) = \exp(-x^2/\lambda^2)$ where $\lambda$ was selected in the same way as CF.} is seen to enjoy the `best of both worlds', behaving like CF in the region of the data and like ZVCV in the tail (\Cref{subfig:gauss_secf}).
Examining the sampling distribution of these estimators through repeated simulation, we observe a remarkable increase in accuracy as a result of post-processing the MCMC output (\Cref{subfig:gauss_boxplot}). Although the total computing time for the 100 repeated simulations increases from approximately 0.02 seconds for vanilla Monte Carlo integration to 0.11 seconds for ZVCV, 0.17 seconds for CF and 0.19 seconds for SECF, all three control variate methods improve upon the vanilla Monte Carlo estimate in terms of the overall efficiency measured by the product of mean square error and computing time.
\end{example*}

\paragraph{Computational Cost}

For many problems the benefit provided by control variates is not justified when the computational cost of implementation is taken into account (see \Cref{tab:CVcomparison}). 
However, when the cost of obtaining MCMC output, or the cost of evaluating $f$ on MCMC output, is sufficiently high then control variates can be a useful tool. 
For borderline cases, \citet{Si2020} demonstrated the use of stochastic gradient descent to speed up the optimisation in Step 3 of the general recipe to select a Monte Carlo estimator. 
A reduced-cost SECF method, based on a low-rank Nystr\"{o}m approximation, was also proposed in \citet{South2020}.

\paragraph{Curse of Dimension}
The gradient-based control variates that we discussed suffer from a curse of dimension, which is most evident in kernel methods like CF. 
However, the regression perspective in \Cref{eq: cv as regression} suggests that, by analogy with high-dimensional regression modelling \citep{buhlmann2011statistics}, it may be possible to construct control variates for functions $f$ whose \textit{effective dimension} is small, despite a high ambient dimension of $\mathcal{X}$.
Additional regularisation can be introduced to this effect \citep{South2018,Zhu2018}, with positive results reported for $d\leq 100$. 
For even larger $d$, it may be sensible to pursue \emph{nonlinear approximation} \citep{devore1998nonlinear}, where the basis $\Phi$ is restricted to allow dependence only on a subset of the parameters \citep[so-called \textit{a priori} regularisation in][]{South2018}.

\subsection{Summary}

This section focused on the application of gradient-based control variates to approximate an integral of interest based on output from MCMC. Applications to other sampling algorithms, such as 
population MCMC \citep{Oates2016}, stochastic gradient Langevin dynamics \citep{Baker2019}, sequential Monte Carlo \citep{South2018} and unbiased MCMC with couplings \citep{South2019discussion}, have also been considered and much of our discussion applies unchanged. 
Applications to estimation of the normalising constant of the posterior have also been considered in the population MCMC and sequential Monte Carlo sampler settings \citep{Oates2016,South2018}. Again, the extension is straightforward and consists of applying the ideas from this section to improve multiple expectations. The \texttt{ZVCV} package \citep{rZVCV} on CRAN provides functions to apply ZVCV and CF to two estimators of the normalising constant.

A current weakness of control variate methodology is that it is under-developed from a theoretical perspective; our focus was on sets of control variates that form linear subspaces of $\mathcal{L}^2(P)$, for which some limited theoretical understanding has been achieved, but more sophisticated sets of control variates have also been empirically considered.
For example, \citet{Zhu2018,Si2020} proposed to use the gradients of neural network for the set $\Phi$. 
A neural network is parameterised by a collection of \textit{weights} and \textit{biases}, which are jointly estimated using stochastic gradient descent applied to a proxy for mean square error, as discussed in \Cref{subsubsec: proxies}. 
These authors found empirically that this approach can lead to improved performance over methods like ZVCV and CF in the high dimensional context.
In light of the anticipated technical complexity required to analyse such sophisticated control variate methods, we expect that empirical assessment will continue to be the primary means through which control variate methodology is developed and assessed.

\section{DISCUSSION}
\label{sec: discuss}

MCMC has become a core part of most graduate programmes in Statistics, due to its effectiveness in enabling Bayesian analyses to be performed.
Perhaps understandably, these programmes focus on the \textit{design} and \textit{validity} of algorithms, emphasising the elegant probabilistic arguments that are often involved.
However, this leaves little or no time to discuss post-processing of MCMC output. 
In fact, our impression is that many professional users of MCMC are also not aware of this aspect, beyond convergence diagnostics and burn-in removal.
Through writing this review, we hope greater attention may be given to this under-appreciated but important practical side of MCMC.
In particular, the topic is receiving considerable attention from computational researchers at the time of writing, and we extend an invitation to the interested reader to explore further into the recent works cited.

\section*{DISCLOSURE STATEMENT}
Aside from being authors of some of the literature that was discussed, the authors are not aware of any affiliations, memberships, funding, or financial holdings that might be perceived as affecting the objectivity of this review. 

\section*{ACKNOWLEDGMENTS}
MR, OT, CJO were supported by the Lloyd's Register Foundation programme on data-centric engineering at the Alan Turing Institute, UK.
MR was supported by the British Heart Foundation - Alan Turing Institute cardiovascular data science award (BHF; SP/18/6/33805) and  by the Wellcome/EPSRC Centre for Medical Engineering (WT203148/Z/16/Z). 
The authors thank Matt Graham, Aki Vehtari, Ioannis Kontoyiannis, Pierre Jacob and an anonymous reviewer for helpful comments.

\bibliographystyle{ar-style1}
\bibliography{bibliography}

\end{document}